\documentclass[english,11pt,aps,prd,a4paper,preprintnumbers,  floatfix,nofootinbib,showpacs,superscriptaddress,  notitlepage]{revtex4-1} 

 \pdfoutput=1
\usepackage[usenames,dvipsnames]{color}  
\usepackage{graphicx}
\usepackage{setspace}
\usepackage{upgreek}
\usepackage{braket}
\usepackage{array}
\usepackage{microtype}
\usepackage{multirow}
\usepackage{tabularx}

\usepackage{bm}
\usepackage{bbm}
\usepackage{mathrsfs}
\usepackage[bbgreekl]{mathbbol}

\usepackage{caption}
\usepackage{stackengine}
\usepackage{subcaption}
\usepackage{amsmath}
\usepackage{amssymb}
\usepackage[colorlinks=true,citecolor=darkred,urlcolor=darkred, pdfborder={0 0 0}]{hyperref}
\usepackage[normalem]{ulem}
\usepackage{xcolor}
\usepackage{graphicx}
\makeatletter
\def\p@subsection{}
\makeatother
\usepackage{xcolor}
\usepackage{float}
\usepackage{placeins}
\usepackage{tabularx}
\usepackage{booktabs} 

\definecolor{darkred}{rgb}{0.6,0,0}

\definecolor{linkcolor}{rgb}{0,0,0.5}

\def\gsim{\raise0.3ex\hbox{$\;>$\kern-0.75em\raise-1.1ex\hbox{$\sim\;$}}}
\def\lsim{\raise0.3ex\hbox{$\;<$\kern-0.75em\raise-1.1ex\hbox{$\sim\;$}}}

\def\beqn#1{\begin{equation}\label{#1}}
\def\eeqn{\end{equation}}

\def\beqa#1{\begin{eqnarray}\label{#1}}
\def\eeqa{\end{eqnarray}}

\usepackage{caption}
\usepackage{subcaption}
\usepackage{adjustbox}

\def\Z2{$\mathcal{Z_2}$}

\usepackage{bbold}
\usepackage{ragged2e}

\newcommand {\ignore}[1]{}

\usepackage{mathrsfs}

\def\321{$\mathrm{SU(3) \otimes SU(2) \otimes U(1)}$ }

\definecolor{matplotlibBrown}{rgb}{0.65, 0.16, 0.16}

\usepackage{orcidlink}

\newcommand{\AddrIISERB}{Department of Physics, Indian Institute of Science Education and Research - Bhopal, \\ 
Bhopal Bypass Road, Bhauri, Bhopal 462066, India}

\newcommand{\AddrCFTP}{%
Departamento de F\'{\i}sica and CFTP, Instituto Superior T\'ecnico, Universidade de Lisboa, Av. Rovisco Pais 1, 1049-001 Lisboa, Portugal}

\begin{document}

\title{\textcolor{BrickRed}{Flavor specific chiral $U(1)_X$ framework for explaining the ATOMKI anomaly}}

\author{Aditya Batra~\orcidlink{0000-0001-6288-5818}}\email{aditya.batra@tecnico.ulisboa.pt}
\affiliation{\AddrCFTP}
\author{F. R. Joaquim~\orcidlink{0000-0002-6711-4606}}\email{filipe.joaquim@tecnico.ulisboa.pt}
\affiliation{\AddrCFTP}
\author{Hemant Prajapati~\orcidlink{0000-0001-5104-9427}}\email{hemant19@iiserb.ac.in}
\affiliation{\AddrIISERB}
\author{Rahul Srivastava~\orcidlink{0000-0001-7023-5727}}\email{rahul@iiserb.ac.in}
\affiliation{\AddrIISERB}

\begin{abstract}
\vspace{0.5cm}
Recent anomalies in nuclear transitions observed by the ATOMKI Collaboration suggest the existence of a new boson with a mass of $\sim 17$ MeV. A theoretically consistent interpretation requires a framework that not only matches the kinematics but also reproduces the observed decay rates while satisfying stringent experimental constraints. Among various possibilities, an axial-vector or mixed vector--axial-vector mediator $Z'$ emerges as the most viable candidate.
However, getting such couplings for a light $Z'$ gauge boson is a highly nontrivial task.
In this work, we construct a gauged chiral, flavor specific $U(1)_X$ extensions of the Standard Model where the associated $Z'$ boson acts as the $17$ MeV particle. By employing a two Higgs doublet framework, we generate the necessary nonvanishing axial-vector couplings while ensuring gauge anomaly cancellation and consistent fermion mass generation. 
Focusing on the $^8\mathrm{Be}$ and $^4\mathrm{He}$ signals, we show that in this model the viable parameter space to resolve the ATOMKI anomalies is also consistent with a diverse set of experimental constraints, including atomic parity violation, beam dump experiments, meson decays, and neutrino nucleus and neutrino electron scatterings. Our results demonstrate that this framework offers a theoretically sound and phenomenologically robust solution to the ATOMKI anomaly.
\end{abstract}

\maketitle  

\section{\label{sec:Intro}Introduction}
In the era of the Large Hadron Collider (LHC), searches for new physics have primarily focused on high-energy phenomena above the TeV scale, motivated largely by the belief that new physics is expected to manifest through new particles which are heavier or near the electroweak scale \cite{Eichten:1983hw,Martin:1997ns,Bajc:2006ia,Preda:2022izo}.  
However, the lack of new discoveries at the LHC has renewed interest in light, weakly coupled new physics at the MeV scale.
Nuclear transitions offer an excellent probe for such light particles, since they are light enough to be produced in nuclear decays \cite{Treiman:1978ge,Freedman:1984sd,Savage:1986ty}.
A nuclear transition occurs when an excited nucleus decays to a lower energy state of the same nucleus. 
In the Standard Model (SM), these transitions are predominantly mediated by electromagnetic interactions and can proceed either via the emission of a real photon $\gamma$ or through internal pair creation (IPC), where a virtual photon $\gamma^*$ subsequently decays into an electron positron pair.
Owing to the large statistics achievable in nuclear experiments, even very weakly coupled new physics can significantly affect the observed transition rates.
This makes nuclear processes powerful probes for rare decays mediated by beyond the Standard Model (BSM) particles, even when their interactions with the SM are extremely feeble.

\begin{figure}[t!]
\begin{center}
\includegraphics[width=0.5\linewidth]{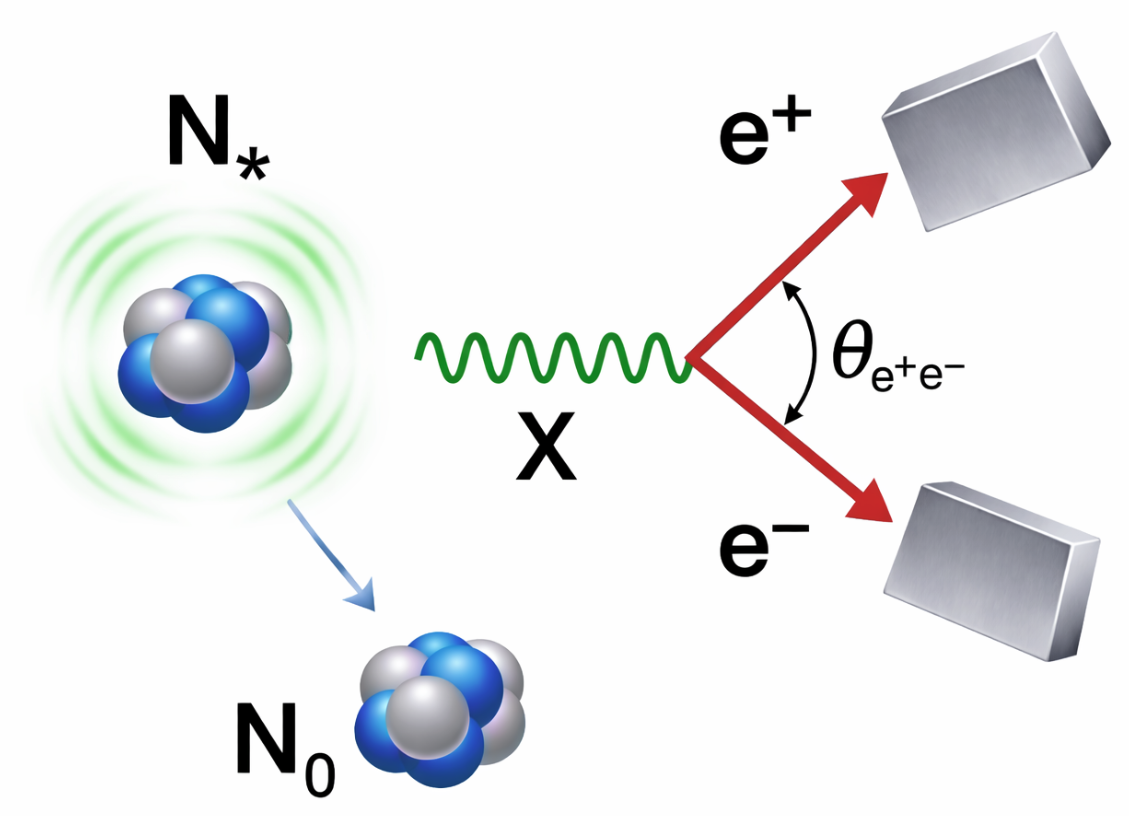}
\end{center}
\caption{Schematic representation of the IPC process. An excited nucleus $N_{*}$ decays to its ground state $N_{0}$ via the emission of a hypothetical light gauge boson $X$, which subsequently decays into an electron positron ($e^+e^-$) pair.}
\label{FiG:IPC}
\end{figure}

In recent years, the ATOMKI Collaboration has reported several anomalies in the IPC decays of excited nuclear states \cite{Krasznahorkay:2015iga,Krasznahorkay:2017gwn,Krasznahorkay_2018,Krasznahorkay:2019lyl,Krasznahorkay:2021joi,refId0,Krasznahorkay:2022pxs,Sas:2022pgm}.
In 2015, the ATOMKI Collaboration reported a $6.8\sigma$ anomaly in the IPC decays of an isoscalar ($18.15$ MeV) M1 transition in $^{8}\mathrm{Be}$ \cite{Krasznahorkay:2015iga,Krasznahorkay_2018}. 
The anomaly corresponds to an excess of events in the decay of the excited beryllium nucleus to its ground state at opening angle $\theta_{e^{+}e^{-}} \sim 140^{\circ}$. The Collaboration also proposed a possible solution in terms of the on shell production of a new boson $X$, which subsequently decays into an electron-positron pair ; see Fig. \ref{FiG:IPC}. 
The best fit mass of this boson is estimated to be $\sim 17$ MeV. 
Later in 2017, the ATOMKI Collaboration also reported an anomaly in the isovector ($17.64$ MeV) M1 transition in $^{8}\mathrm{Be}$ \cite{Krasznahorkay:2017gwn}.
Since then, the ATOMKI Collaboration has also reported anomalies in the 
the decays of excited $^{4}\mathrm{He}$~\cite{Krasznahorkay:2019lyl,Krasznahorkay:2021joi,refId0} ($\theta_{e^{+}e^{-}} \sim 115^{\circ}$) and, more recently, in $^{12}\mathrm{C}$ ($\theta_{e^{+}e^{-}} \sim 150^{\circ}-160^{\circ}$) \cite{Krasznahorkay:2022pxs} nuclei, all with significances above $5\sigma$. 
Kinematically, the anomalies are consistent with the production of a new boson of mass $\sim 17$~MeV that subsequently decays into an electron positron pair \cite{Denton:2023gat}.
For a long time, no independent probes of these anomalies were available. In 2023, the MEG II Collaboration released the results of their analysis based on four weeks of data \cite{Anh:2024req,MEGII:2024urz} and reported no significant observation above the expected background. Still, their results remain compatible with ATOMKI within $\sim 1.5\sigma$ due to limited statistics \cite{Barducci:2025hpg}. 
More recently, in 2025, the PADME experiment provided a completely independent search by employing a positron beam incident on a fixed target to probe for a hypothetical particle with a mass of approximately 17 MeV \cite{Bossi:2025ptv,Arias-Aragon:2025wdt}.
A $\sim 2\sigma$ deviation from the null hypothesis was observed at $\sqrt{s} \approx 16.90$ MeV, with the location of this excess consistent with the average mass favored by the ATOMKI experiments.
This observation has once again sparked interest in the so called $X17$ boson.

Kinematically, the ATOMKI anomaly is consistent with the production of a new boson of mass around $17$~MeV. However, a compelling new physics interpretation requires more than mere kinematic consistency with the best fit mass. A description in terms of a new particle demands dynamical consistency as well: the rates of the anomalous decays must be reproducible using the same set of interaction strengths between the new boson $X$ and quarks.
Many theoretical studies have been carried out to analyze the nature of this hypothetical $X$ boson \cite{Feng:2016jff,Feng:2016ysn,Feng:2020mbt,Barducci:2022lqd,Hostert:2023tkg,Mommers:2024qzy,Fieg:2026zkg}. 
Following the $^{12}\mathrm{C}$ result, the authors of Ref. \cite{Barducci:2022lqd} examined the possible theoretical interpretations and the nature of the new boson.
A pure scalar mediator cannot explain the $^8$Be anomaly, while a pure pseudoscalar cannot account for the $^{12}$C anomaly. 
In the case of a pure vector mediator, it has been 
demonstrated that such a particle is unable to simultaneously 
accommodate all the observed ATOMKI anomalies.
Furthermore, 
Ref.~\cite{Hostert:2023tkg} showed that the vector boson interpretation of the ATOMKI signal is ruled out by the limits imposed by searches 
for visible resonances in three track pion decays 
($\pi^+ \to e^+ \nu_e e^+ e^-$) at the SINDRUM-I experiment.
Results from Ref. \cite{Barducci:2022lqd,Hostert:2023tkg}  show that the new $X$ boson being an axial-vector (or mixed vector--axial-vector) boson is the most promising candidate, whereas other spin/parity assignments are incompatible with a combined explanation of the $^{8}\mathrm{Be}$, $^{4}\mathrm{He}$, and $^{12}\mathrm{C}$ anomalies. In such a case, typical axial coupling to quarks in the range $\sim 10^{-4} - 10^{-3}$ is required to reproduce the observed excess that simultaneously explains $^{8}\mathrm{Be}$ and $^{4}\mathrm{He}$ anomalies. However, the axial-vector interpretation for the $^{12}\mathrm{C}$ case is subject to more significant uncertainties, largely due to the lack of available data on the $^{12}\mathrm{C}$ transition matrix elements.
Recently, Ref.~\cite{Mommers:2024qzy} computed the nuclear matrix elements for the 17.23 MeV excited state in $^{12}$C, assuming it is fully described by a $2s_{1/2} \, 1p_{3/2}^{-1}$ particle-hole excitation relative to the ground state. They found that explaining the $^{12}$C anomaly at the $1\sigma$ level requires at least one nucleon coupling of order $\mathcal{O}(10^{-2})$. Such large couplings are strongly constrained by other experiments~\cite{Fieg:2026zkg,Mommers:2024qzy}. However, at significance levels $\gtrsim 2\sigma$, the theoretical and experimental uncertainties are sufficiently large that no definitive conclusion can be drawn. Moreover, this calculation predicts a photon decay width of 251 eV, substantially larger than the experimentally measured value of 44 eV, indicating the limitations of this simple shell model approximation. Nevertheless, new experimental measurements of the transition and/or improved nuclear modeling could clarify the situation.

Many efforts have been carried out to build UV complete models to explain these anomalies ~\cite{Feng:2016ysn, Feng:2016jff,Ellwanger:2016wfe,Pulice:2019xel,Feng:2020mbt,Nomura:2020kcw,Ferreira:2023buj,Abdallah:2024uby}.
A spin-1 mediator with axial-vector couplings (or mixed vector and axial-vector couplings) comes out to be the most viable candidate for explaining the ATOMKI 
anomaly \cite{Barducci:2022lqd,Hostert:2023tkg}. Such a framework has been explored in Refs.~\cite{Kahn:2016vjr, Fayet:2016nyc, 
Gu:2016ege, DelleRose:2017xil, DelleRose:2018eic,Abbaslu:2024hep,Dutta:2026dnf}.
However, constructing a fully UV complete $\mathrm{SM} \otimes U(1)_X$ model that simultaneously generates axial-vector (or mixed) couplings of the $Z'$ to SM fermions while correctly reproducing all SM fermion masses, mixings, and couplings to the SM bosons is nontrivial.
A chiral $U(1)_X$ symmetry, in which left- and right-handed fermions carry distinct charges, provides a viable mechanism to generate these mixed vector and axial-vector interactions~
\cite{Appelquist:2002mw,Montero:2007cd,Ma:2014qra,Oda:2015gna,Das:2016zue,Jana:2019mez,Mandal:2023oyh,Prajapati:2024wuu,Prajapati:2026tfv}.
This chiral structure, however, introduces two primary constraints on $U(1)_X$ charges of fermions. First, the fermion $U(1)_X$ charges must satisfy 
a set of gauge anomaly cancellation conditions to preserve the 
unitarity and renormalizability of the 
theory~\cite{Adler:1969gk, Bardeen:1969md, Bell:1969ts, 
Delbourgo:1972xb, Witten:1982fp, Alvarez-Gaume:1983ihn}. Second, 
the Higgs must be charged under $U(1)_X$ to allow for 
gauge invariant Yukawa interactions that generate the SM fermion masses. Together, these requirements impose stringent constraints on the allowed $U(1)_X$ charge assignments for SM fermions~\cite{Prajapati:2024wuu,Prajapati:2026tfv}.

The simplest implementation 
of such chiral symmetry is a flavor universal construction, where all three generations of SM fermions are assigned identical $U(1)_X$ 
charges. 
This approach is particularly advantageous, as it automatically avoids $Z'$ mediated flavor changing neutral currents (FCNCs) at the tree level.
However, in such constructions, the axial-vector couplings for a light 
$Z'$ vanishes at the leading order in the gauge coupling due to the gauge invariance requirements of the Yukawa 
interactions~\cite{Kahn:2016vjr, DelleRose:2018eic,Prajapati:2026tfv}. 
This suppression is a generic feature of any $U(1)_X$ extension of the SM featuring a 
single Higgs doublet whose Yukawa couplings respect the $U(1)_X$ gauge symmetry~\cite{Kahn:2016vjr,Prajapati:2026tfv}.
This suppression of axial-vector couplings is effectively avoided in two Higgs doublet models with flavor specific $U(1)_X$ charge assignments. Such constructions provide a natural mechanism for generating nonvanishing axial-vector couplings, as the additional Higgs doublet can participate in writing the Yukawa interactions responsible for generating fermion masses ~\cite{Kahn:2016vjr,DelleRose:2017xil,Allanach:2018vjg,Prajapati:2026tfv}.

In this work, to address the ATOMKI anomalies, we consider a gauged chiral flavor specific $U(1)_X$ extension of the SM. The associated neutral gauge boson $Z'$ serves as the hypothesized $\sim 17$~MeV $X$ particle. 
Given the ongoing uncertainty in the nuclear matrix elements for the $^{12}$C anomaly, we focus on models that accommodate the signals in $^8$Be (18.15), $^8$Be (17.64), and $^4$He, without relying on the calculations presented in Ref.~\cite{Mommers:2024qzy}.
The remainder of this paper is organized as follows. In Sec.~\ref{Sec:challenges}, we provide an overview of the current experimental and theoretical status of the ATOMKI anomaly, while in Sec.~\ref{Sec:axialcoupling} we discuss how the required axial-vector couplings between quarks and the $Z'$ can be generated within a two Higgs doublet framework.  
Section~\ref{Sec:Exp_Constraints} 
reviews the various experimental constraints relevant to these couplings.
In Sec.~\ref{Sec:Model}, we construct an explicit, anomaly-free model that realizes the fermion coupling structure discussed previously, where the 
$U(1)_X$ charges are chosen to yield axial-vector couplings of the order 
$10^{-4}-10^{-3}$.
Our main results are collected in Sec.~\ref{Sec:results}, 
where we present the allowed parameter space that explains the ATOMKI anomalies while satisfying constraints from the atomic parity violation, 
beam-dump experiments, anomalous magnetic moments, meson decays, and 
neutrino scattering. Finally, in Sec.~\ref{sec:conclusion}, we summarize our findings and provide concluding remarks.
\section{Why is it so challenging to explain the ATOMKI anomalies?}
\label{Sec:challenges}
In the ATOMKI experiment, a proton beam collides with a nucleus $A$ at rest to produce an excited nuclear state $N_{*}$, via the process $p + A \to N_{*}$. The IPC decay of $N_{*}$ back to the ground state, $N_{*} \to N_{0} + e^+ e^-$, is then analyzed.
In the past decade, the ATOMKI Collaboration has studied IPC decays in excited nuclear states $^{8}$Be (18.15), $^{8}$Be (17.64), $^{4}$He (20.21--21.01), and $^{12}$C (17.23), having reported anomalies in all these decays, each with a local significance exceeding $5\sigma$. Their measurements are summarized in Table~\ref{Tab:atomki_Exp_results}, where one can see that the anomaly corresponds to an excess of events at a large opening angle $\theta_{e^+e^-}$, for the corresponding excited nuclear state $N_{*}$.
\begin{table}[t!]
\centering
\renewcommand{\arraystretch}{1.6} 
\setlength{\tabcolsep}{2.6pt} 
\small
\begin{tabularx}{\textwidth}{| c c | c  | c | c |}
\hline
 & & \multicolumn{3}{c|}{\textbf{ATOMKI Measurements}} \\ 
\cline{3-5}
$N_{*}$ & $J_{*}^{P_{*}} \to J_{0}^{P_{0}}$ & Excess at $\theta_{e^{+}e^{-}}$ & $M_X$ (MeV) & \centering Decay rates \tabularnewline 
\hline
$^{8}$Be (18.15) & $1^{+} \to 0^{+}$ & $\sim 140^{\circ}$ & $16.7 \pm 0.35 \pm 0.50$ & $B_{X}=(6 \pm 1)\times 10^{-6}$ \cite{Krasznahorkay:2015iga,Krasznahorkay_2018} \\ 
\hline
$^{8}$Be (17.64) & $1^{+} \to 0^{+}$ & $\sim 150^{\circ}$& $17.0 \pm 0.50 \pm 0.50$ & $B_{X}=(2 \pm 2)\times 10^{-6}$ \cite{Krasznahorkay:2017gwn}\\ 
\hline
$^{4}$He (20.21), $^{4}$He (21.01) & $0^{+},0^{-} \to 0^{+}$ & $\sim 115^{\circ}$& $16.98 \pm 0.16 \pm 0.20$&$\sigma_{X}/\sigma_{\text{E0}}=0.2$ ~\cite{Krasznahorkay:2019lyl,Krasznahorkay:2021joi,refId0}\\ 
\hline
$^{12}$C (17.23) & $1^{-} \to 0^{+}$ & $\sim (150^{\circ}-160^{\circ})$& $17.03 \pm 0.11 \pm 0.20$&$B_{X}=(3.6 \pm 0.3)\times 10^{-6}$ \cite{Krasznahorkay:2022pxs}\\
\hline
\end{tabularx}
\caption{Summary of ATOMKI experimental results for the IPC transitions $N_{*} \to N_{0} + e^{+}e^{-}$. The table lists the spin parity assignments ($J^{P}$), the opening angle $\theta_{e^{+}e^{-}}$ of the observed excess, the reconstructed mass of $X$ boson $M_X$, and the ratio of the partial decay width of $X$ to that of the photon or cross section ratios for each anomaly. See text for more details. }
\label{Tab:atomki_Exp_results}
\end{table}
These observations can be interpreted as the on shell production of a new light, short lived neutral boson $X$ in the decay $N_{*} \to N_0 X$, which subsequently decays into an electron-positron pair, $X \to e^+ e^-$. Kinematically, the observed bumps in the $\theta_{e^+e^-}$ and $e^+e^-$ invariant mass distributions are consistent with the production of a single new boson $X$ with a best fit mass of approximately $17$~MeV.
However, a compelling new physics interpretation requires more than mere kinematic consistency with the best fit mass. A description in terms of a new particle demands dynamical consistency as well: the rates of the anomalous decays must be reproducible using the same set of interaction strengths between the new boson X and quarks. 
The normalized IPC decay rates relative to the photon emission,
\begin{equation}
    B_X = \frac{\Gamma(N_* \to N + X)}{\Gamma(N_* \to N + \gamma)} \times \text{BR}(X \to e^+ e^-) \,,
\end{equation}
%
required to explain the $^{8}\text{Be}$ and $^{12}\text{C}$ anomalies, are presented in the final column of Table~\ref{Tab:atomki_Exp_results}. 
In the case of Helium, the proton beam energy is tuned to lie between two nearby resonances of $^{4}\mathrm{He}$ at $20.21$ and $21.01~\mathrm{MeV}$. 
In this setup, both excited states are populated and can contribute to the anomalous signal. Therefore, the relevant observable for the $^{4}\mathrm{He}$ transition is the total cross section, defined as the 
sum of the contributions from both states, $\sigma_{X}$, normalized by the SM E0 transition in $^{4}\mathrm{He}$ $(20.21)$ \footnote{For $^{4}\mathrm{He}(21.01)$, the E0 transition is forbidden due to parity conservation in electromagnetic interactions.}. 

\begin{figure}[t!]
\begin{center}
\includegraphics[width=0.49\linewidth]{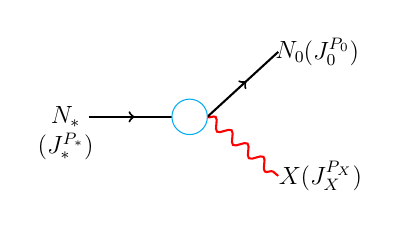}
\end{center}
\caption{ Schematic representation of the on shell emission of a hypothesized $X$ boson during the nuclear transition $N_{*} \to N_{0} + X$. The variables $J_{*}^{P_*}$, $J_{0}^{P_0}$, and $J_{X}^{P_X}$ denote the spin parity assignments of the excited state, ground state, and the new boson, respectively.}
\label{fig:Spin}
\end{figure}

The spin parity quantum numbers of the excited nuclear state $N_*$ and the ground state $N_0$, denoted by $J_*^{P_*}$ and $J_0^{P_0}$, respectively, are shown in Table~\ref{Tab:atomki_Exp_results} for the several studied transitions. As for the new boson $X$, the possible spin and parity assignments are $J_X^{P_X}=0^+$ (scalar), $J_X^{P_X}=0^-$ (pseudoscalar), $J_X^{P_X}=1^-$ (vector) or $J_X^{P_X}=1^+$ (axial-vector), see Fig.~\ref{fig:Spin}.
Assuming parity conservation in the transition $N_* \to N_{0} + X$, provides a useful constraint on the spin parity assignment of the hypothetical $X$ boson. Specifically, since $J_{*}^{P_{*}}$ and $J_{0}^{P_{0}}$ are well established, the allowed assignments for $J_{X}^{P_X}$ are strictly governed by conservation laws. Namely, angular momentum and parity conservation yields
\begin{equation}\label{Eq:Spin_parity}
    J_{*} = L \oplus J_{X} ,~~~~P_{*} = (-1)^{L}P_{X}\,,
\end{equation}
where $L$ is the relative angular momentum between $N_{0}$ and $X$. Since $J_{*}^{P_{*}}$ differs among the excited states, the allowed $J_{X}^{P_X}$ assignments, inferred from Eq.~\eqref{Eq:Spin_parity}, will depend on the specific state under consideration. This is apparent from Table~\ref{Tab:possibility_for_X} where we present the various spin parity assignments for the $X$ boson that are permitted by angular momentum and parity conservation for different $N_{*}$.
\begin{table}[ht]
\centering
\renewcommand{\arraystretch}{1.6}
\setlength{\tabcolsep}{10pt}
\begin{tabular}{|c c|c|c|c|c|}
\hline
\multicolumn{2}{|c|}{} & \multicolumn{4}{c|}{\textbf{Possibilities for } $X$} \\
\cline{3-6}
$N_*$ & $J_*^{P_*} \to J_0^{P_0}$ & Scalar & Pseudoscalar & Vector & Axial-Vector \\
\hline
$^{8}$Be (18.15), $^{8}$Be (17.64) & $1^+ \to 0^+$ & \textcolor{red}{\textbf{×}} & $L=1$ & $L=1$ & $L=0,2$ \\
\hline
$^{4}$He (20.21) & $0^+ \to 0^+$ & $L=0$ & \textcolor{red}{\textbf{×}} & $L=1$ & \textcolor{red}{\textbf{×}} \\
\hline
$^{4}$He (21.01) & $0^- \to 0^+$ & \textcolor{red}{\textbf{×}} & $L=0$ & \textcolor{red}{\textbf{×}} & $L=1$ \\
\hline
$^{12}$C (17.23) & $1^- \to 0^+$ & $L=1$ & \textcolor{red}{\textbf{×}} & $L=0,2$ & $L=1$ \\
\hline
\end{tabular}
\caption{Permissible spin parity assignments (Scalar, Pseudoscalar, Vector, and Axial-Vector) for the hypothesized $X$ boson across various nuclear transitions. The entries denote the required orbital angular momentum $L$ between the $X$ boson and the ground state $N_0$ necessary to satisfy the conservation laws in Eq.~\eqref{Eq:Spin_parity}. Red marks (\textcolor{red}{\textbf{×}}) indicate assignments that are forbidden by parity and angular momentum conservation for the given transition.}
\label{Tab:possibility_for_X}
\end{table}
%
Several conclusions can be inferred from Table~\ref{Tab:possibility_for_X}, namely:
\begin{itemize}
    \item A pure scalar mediator cannot explain the $^8$Be anomaly.
    \item A pure pseudoscalar can account for the $^{8}\text{Be}$ anomaly and also the $^{4}\text{He}$ one, provided that in the latter case the transitions are dominated by the $^{4}$He (21.01) excited state. However, pure pseudoscalar fails to explain the anomaly observed in $^{12}\text{C}$.
    \item A vector or axial-vector $X$ boson can possibly explain all three anomalies, but only one of the two $^{4}\mathrm{He}$ resonant states can contribute to the anomalous signal.
\end{itemize}

Hence, the spin parity assignments suggest that a vector or axial-vector mediator can explain all the anomalies simultaneously. However, any viable new physics explanation must also reproduce the observed decay rates listed in Table~\ref{Tab:atomki_Exp_results} for all anomalous channels with a consistent set of $X$ couplings to quarks. Obviously, these couplings must remain compatible with various experimental constraints, including atomic parity violation, beam dump experiments, anomalous magnetic moments, meson decays, neutrino scattering, and collider searches. A comprehensive discussion on these constraints and their implications is provided in Sec.~\ref{Sec:Exp_Constraints}.

At this point, the following conclusions can be drawn from the existing literature:
\begin{itemize}
    \item Based on spin parity considerations alone, a pure scalar mediator cannot explain the $^{8}\mathrm{Be}$ anomaly, while a pure pseudoscalar cannot account for the $^{12}\mathrm{C}$ anomaly.
    \item In principle, a mixed scalar-pseudoscalar scenario could explain all 
    anomalies simultaneously. 
   As discussed in Ref.~\cite{Barducci:2022lqd}, 
  if only the $^{8}\text{Be}$ (18.15), $^{4}\text{He}$, and 
  $^{12}\text{C}$ excited states are considered, a consistent 
  interpretation is possible, although the allowed parameter space is severely restricted by the SINDRUM-I bounds. 
  However, once the 
  $^{8}\text{Be}$ (17.64) state is included alongside the $^{8}\text{Be}$ (18.15) 
  resonance, a combined explanation of the $^{8}\text{Be}$ and 
  $^{4}\text{He}$ anomalies is not possible.
 \item For the pure vector case, a combined explanation of the 
$^{8}\text{Be}$, $^{4}\text{He}$, and $^{12}\text{C}$ signals is in 
conflict with protophobic constraints derived from neutral pion decays ($\pi^{0} \to \gamma X$)~\cite{Barducci:2022lqd}.
Additionally, Ref.~\cite{Hostert:2023tkg} 
derived constraints on the couplings of light vector particles using 
charged pion decays ($\pi^{+} \to e^{+} \nu_{e} X$) from the 
SINDRUM-I experiment. These results demonstrate that, for a pure vector mediator, the charged pion decay limits impose significant 
tension even between the initial $^{8}\text{Be}$ and $^{4}\text{He}$ 
anomalies. Consequently, the pure vector boson scenario is 
disfavored by current pion decay constraints.
 \item A pure axial-vector scenario can explain $^{8}\text{Be}$ (18.15), $^{8}\text{Be}$ (17.64) and $^{4}$He anomalies~\cite{Barducci:2022lqd}. Regarding the $^{12}$C anomaly, the axial vector case is subject to more significant uncertainties, and the lack of data on the $^{12}{\rm C}$ transition matrix elements.
As mentioned earlier, the only available computation~\cite{Mommers:2024qzy} requires at least one nucleon coupling of order $\mathcal{O}(10^{-2})$ at the $1\sigma$ level. Such large couplings are strongly constrained by other experiments~\cite{Fieg:2026zkg,Mommers:2024qzy}. Furthermore, at the $\gtrsim 2\sigma$ level, the theoretical and experimental uncertainties are sufficiently large that no definitive conclusion can be drawn. Moreover, this calculation predicts a photon decay width of 251 eV, substantially larger than the experimentally measured value of 44 eV, indicating the limitations of this simple shell model approximation.
Nevertheless, new experimental measurements of the transition and/or improved nuclear modeling could clarify the situation.
\end{itemize}

In conclusion, the axial-vector (or mixed vector--axial-vector) mediator emerges as the most promising candidate for a simultaneous explanation of at least the $^{8}\text{Be}$ 
and $^{4}\text{He}$ anomalies 
\footnote{In scenarios where both vector and axial-vector couplings are of comparable magnitude, the axial-vector contribution typically dominates the transition rates~\cite{Barducci:2022lqd}.}.
In this case, typical axial coupling to quarks in the range $\sim 10^{-4} - 10^{-3}$ is required to reproduce the observed excess while satisfying other experimental bounds. Regarding the $^{12}\text{C}$ signal, we emphasize that a definitive interpretation requires a more comprehensive nuclear matrix calculations. Given that such calculations are currently unavailable, further definitive comments on this channel cannot be made at this time.

\section{Generating an axial vector coupling in $U(1)_X$ models}
\label{Sec:axialcoupling}
In this work, we build a flavor specific $U(1)_{X}$ model that can generate sizable axial couplings of the new boson $Z'$ with SM fermions. The present section is devoted to discussing how to construct an axial vector coupling between $Z'$ and SM fermions in $U(1)_X$ gauged models. \footnote{An explicit model realizing this and evading all experimental constraints is presented in Sec.~\ref{Sec:Model}.}
The interaction between gauge bosons and fermions originates from the kinetic term of the fermion fields in the Lagrangian,
\begin{equation}\label{Eq:Kinetic_fermion}
\mathcal{L}_{\rm kinetic} = \sum_{\psi} i \bar{\psi} \gamma^{\mu} D_{\mu} \psi,
\end{equation}
where the summation runs over all fermion fields.
For the 
$\text{SM} \otimes U(1)_X$ gauge extension, the full covariant 
derivative is defined as,
\begin{align}\label{Eq:Covariant_Derivative_general}
D_{\mu}=\partial_{\mu}+ig_{s}T_{s}^{a}G^{a}_{\mu}+igT^{a}W^{a}_{\mu}+ig'\frac{Y}{2} B_{\mu} + iXg_{x}C_{\mu}\,.
\end{align}
In this context, $g_s$, $g$, $g'$   and $g_x$ denote the gauge couplings associated with $SU(3)_c$, $SU(2)_L$, $U(1)_Y$, and $U(1)_X$, respectively. The $U(1)_{Y,X}$ charges will be denoted by $Y$ and $X$, and the generators of $SU(3)_c$ and $SU(2)_L$ are $T_g^a$ and $T^a$, respectively, with $T^a = \sigma^a / 2$, where $\sigma^a$ are the Pauli matrices. The gauge fields defined in Eq.~\eqref{Eq:Covariant_Derivative_general} do not correspond to physical mass eigenstates. These are obtained through a transformation that accounts for mixing between the gauge bosons. This mixing can arise from two distinct sources: kinetic mixing, which originates from the mixing between the kinetic terms of the $U(1)_Y$ and $U(1)_X$ gauge bosons, or mass mixing, which occurs when the vacuum expectation values (VEVs) of the scalar sector generate nondiagonal terms in the gauge boson mass matrix.
\subsection{Kinetic mixing}
We begin with the kinetic mixing between the SM hypercharge $U(1)_Y$ and an additional $U(1)_X$ gauge symmetry.
The relevant part of the Lagrangian, before field redefinitions, is given as,
\begin{align}
\mathcal{L}_{\rm Kinetic} = -\frac{1}{4}B^{\mu\nu}B_{\mu\nu} - \frac{1}{4}C^{\mu\nu}C_{\mu\nu} - \frac{\kappa}{2} B^{\mu\nu}C_{\mu\nu},
\label{eq:kinetic-mixing}
\end{align}
where $B_{\mu\nu}$ and $C_{\mu\nu}$ are the $U(1)_Y$ and $U(1)_X$ field strength tensors, respectively. The parameter $\kappa$ characterizes the kinetic mixing between $U(1)_Y$ and $U(1)_X$. Requiring positive definite kinetic energy imposes the condition 
$|\kappa| < 1$.
One can diagonalize the kinetic mixing term as follows,
\begin{align}
 \begin{pmatrix}
  B_{\mu} \\
  C_{\mu}  \\ 
 \end{pmatrix}
=
\begin{pmatrix}
 1 & -\frac{\kappa}{\sqrt{1-\kappa^2}}  \\
 0 & \frac{1}{\sqrt{1-\kappa^2}}\\
\end{pmatrix}
\begin{pmatrix}
 \tilde{B_{\mu}} \\
 \tilde{C_{\mu}} \\
\end{pmatrix}.
\end{align}
With these definitions, the covariant derivative in 
Eq.~\eqref{Eq:Covariant_Derivative_general} can be expressed as:
\begin{align}\label{Eq:Covariant_Derivative}
D_{\mu}=\partial_{\mu}+ig_{s}T_{s}^{a}G^{a}_{\mu}+igT^{a}W^{a}_{\mu}+ig'\frac{Y}{2} \tilde{B}_{\mu} + iX\tilde{g}_{x}\tilde{C}_{\mu}\,,
\end{align}
where $\tilde{g}_X$ represents the effective $U(1)_X$ gauge coupling.
In the limit of small kinetic mixing, this effective gauge coupling is given by,
\begin{equation}
    \tilde{g}_{x} = g_x + \frac{Y}{X}\tilde{g}\,,~~~~\tilde{g}=-\frac{\kappa g'}{2}\,.
\end{equation}
\subsection{Mass mixing}
Mass mixing typically arises following spontaneous symmetry breaking (SSB), where the VEVs of the scalar fields generate off diagonal terms in the gauge boson mass matrix. Since the gauge boson spectrum depends solely on the scalar sector, we adopt a general framework discussed in Ref. \cite{Prajapati:2026tfv} to ensure our results are widely applicable. 
This framework consists of two Higgs doublets, $\Phi$ and $\varphi$, and an arbitrary number of SM singlet scalars $\chi_i$.
A second Higgs doublet is required when the $U(1)_X$ charges are flavor specific to generate masses for all SM fermions (see Sec.~\ref{Sec:Model}). 
Meanwhile, the singlet scalars $\chi_i$ allow the $Z'$ mass to vary independently of the electroweak scale and may also contribute to neutrino mass generation and mixing.
To keep the discussion as model independent as possible, in this section we also consider $U(1)_X$ charges of fermions as free parameters
\footnote{We initially also treat the scalar $U(1)_X$ charges as free parameters. However, constraints on their charges arising from Yukawa interactions are discussed later in this section.}.
Constraints on these charge assignments arising from gauge anomaly cancellation are presented in Sec.~\ref{Sec:Model}.

The mass spectrum of gauge bosons arises from the expansion of the kinetic terms of the scalar fields after SSB of both the electroweak and $ U(1)_X$ symmetries. The kinetic terms of the scalars are written as,
\begin{equation}\label{Eq:Kinetic_term_scalar}
(D_{\mu}\Phi)^{\dagger}D^{\mu}\Phi+ (D_{\mu}\varphi)^{\dagger}D^{\mu}\varphi+(D_{\mu}\chi_{i})^{\dagger}D^{\mu}\chi_{i}~.
\end{equation}
Both the Higgs doublets, $\Phi$ and $\varphi$, and the SM singlet scalars, $\chi_i$, acquire  VEVs, thereby breaking the electroweak and $U(1)_X$ symmetries. The corresponding VEVs are given by,
\begin{equation}\label{VEV}
\langle \Phi \rangle = \frac{1}{\sqrt{2}}\begin{bmatrix}
0 \\
v_{\Phi}
\end{bmatrix},~~~~\langle \varphi \rangle = \frac{1}{\sqrt{2}}\begin{bmatrix}
0 \\
v_{\varphi}
\end{bmatrix},~~~~ \langle \chi_{i} \rangle = \frac{v_{i}}{\sqrt{2}}~.
\end{equation} 
Substituting the covariant derivative and the scalar fields, as defined in Eqs.~\eqref{Eq:Covariant_Derivative} and \eqref{VEV}, into Eq.~\eqref{Eq:Kinetic_term_scalar}, we obtain the gauge boson mass matrices.
The absence of mixing in the charged sector implies that the $W$ boson mass is identical to the SM expression, $M_W^2 = g^2 v^2 / 4$, where $v = \sqrt{v_\Phi^2 + v_\varphi^2} \approx 246.22~\text{GeV}$.
The neutral gauge bosons, however, mix with one another, and the corresponding mass matrix in the basis $(\tilde{B}_{\mu}, W_{\mu}^{3}, \tilde{C}_{\mu})$ takes the form,
\begin{equation}\label{Eq:Gauge_Boson_Mass_mat}
\mathcal{M}^2_{_{V}}= \frac{v^{2}}{4}\begin{pmatrix}
g'^{2} & -gg' & 2qg'g_x\\
-gg'   &  g^{2} & -2qgg_x\\
2qg'g_x & -2qgg_x & 4u^{2}g_x^{2}
\end{pmatrix}\,.
\end{equation}
Where,
\begin{eqnarray}\label{Eq:Exact_Mass_mat_para}
&& v^{2} = v_{\Phi}^{2} + v_{\varphi}^{2} \approx (246.22~ \text{GeV})^2\,,~~~~ q = X_{\Phi} +\frac{\tilde{g}}{g_x} + \frac{v_{\varphi}^{2}}{v^{2}} (X_{\varphi} - X_{\Phi}) \,, \\  && u^{2}=\left(X_{\Phi} + \frac{\tilde{g}}{g_x} \right)^{2} + \frac{u_{\chi}^{2}}{v^{2}} + \frac{v_{\varphi}^{2}}{v^{2}} \left[X_{\varphi}^{2} - X_{\Phi}^{2} +2 \frac{\tilde{g}}{g_x} (X_{\varphi}-X_{\Phi}) \right]\;,\; u_{\chi}=\sqrt{\sum_{i}(X^{^{2}}_{\chi_{_{i}}}v_{i}^{2})}\,.
\end{eqnarray}  
Here, $X_{\chi_i}$ denotes the $U(1)_X$ charge of the singlet scalar $\chi_i$. Similarly, $X_\Phi$ and $X_\varphi$ denote the $U(1)_X$ charges of the Higgs doublets $\Phi$ and $\varphi$, respectively.
Notice that, in the limit $v_{\varphi}^{2} \ll v_{\Phi}^{2}$, these parameters can be approximated as follows,
\begin{equation}
    q \simeq X_{\Phi} + \frac{\tilde{g}}{g_x},~~~ u^{2} \simeq q^{2}+\frac{u_{\chi}^{2}}{v^{2}}\,.
\end{equation}
For the remainder of this work, we restrict our analysis to this limit, corresponding effectively to the alignment regime \cite{Branco:2011iw}. 
This limit ensures that the lightest CP-even scalar has SM-like couplings to the gauge bosons \cite{Branco:2011iw}.

The mass matrix in Eq.~\eqref{Eq:Gauge_Boson_Mass_mat} can be diagonalized with orthogonal matrix $\mathcal{O}(\alpha)$, where $\alpha$ is the $Z$--$Z'$ mixing angle.  
The gauge eigenstates $(\tilde{B}^{\mu}, W_{3}^{\mu}, \tilde{C}^{\mu})$ and the mass eigenstates $(A^\mu, Z^\mu, Z^{\prime \mu})$ are related by,
\begin{equation}
\label{unitary matrix}
\begin{bmatrix}
A^{\mu} \\
Z^{\mu} \\
Z^{\prime \mu}
\end{bmatrix} = \mathcal{O}(\alpha) \begin{bmatrix}
\tilde{B}^{\mu} \\
W_{3}^{\mu}\\
\tilde{C}^{\mu}
\end{bmatrix}.
\end{equation}
The resulting diagonal mass matrix is
\begin{equation}\label{Eq:Gauge_and_mass_eigen_states}
\mathcal{M}^2_{\text{diag}} = \mathcal{O}(\alpha)\, \mathcal{M}^2_{V}\, \mathcal{O}^{\dagger}(\alpha)\,,
\end{equation}
and the rotation matrix is defined as,
\begin{equation}\label{Eq:O_alpha_Matrix}
    \mathcal{O}(\alpha) = \begin{bmatrix}
\cos\theta_{W} &~ \sin\theta_{W} &~0\\
-\cos\alpha \sin\theta_{W} & \cos\alpha \cos\theta_{W}
&~ -\sin \alpha\\
-\sin\alpha \sin\theta_{W} &~  \sin\alpha\cos\theta_{W} &~ \cos\alpha 
\end{bmatrix}\,,
\end{equation}
with $\theta_{W}$ denoting the weak mixing angle. Following this rotation, the massless eigenstate is identified as the photon, while the remaining two neutral gauge boson masses read,
\begin{equation}\label{Eq:Gauge_boson_mass}
\begin{aligned}
M_{Z'}^{2}= \frac{v^{2}}{8}\left(A_{0}-\sqrt{B_{0}^{2}+C_{0}^{2}}\right),&&&&&&\,M_{Z}^{2}=\frac{v^{2}}{8}\left(A_{0}+\sqrt{B_{0}^{2}+C_{0}^{2}}\right)\,,
\end{aligned}
\end{equation}
where,
\begin{equation}\label{Eq:A0_B0_C0}
\begin{aligned}
      A_{0}= g^{2}+{g'}^{2}+4u^{2}g_x^{2} \,,~ ~   B_{0}= 4 q g_x\sqrt{g^{2}+{g'}^{2}} \,,~ ~
      C_{0}=  g^{2} + {g'}^{2} - 4u^{2}g_x^{2}\,.
      \end{aligned}
\end{equation}
The $u_{\chi}$ parameter defined in Eq.~\eqref{Eq:Exact_Mass_mat_para} can be expressed as a function of $M_{Z'}$ and $g_x$,
\begin{equation}
\label{Eq:VEV_Paramter_ux}
u_{\chi} =  \frac{M_{Z'}}{g_x}\sqrt{   \frac{ [(M_{Z}^{\text{SM}})^{2} + v^{2}g_x^{2}q^{2}] - M_{Z'}^{2}}{(M_{Z}^{\text{SM}})^{2}  - M_{Z'}^{2} }  }\,. 
\end{equation}
Here, $M_{Z}^{\text{SM}}$ is the mass of $Z$
boson in SM at tree level, defined as $M_{Z}^{\text{SM}} = v\sqrt{g^{2}+g'^{2}}/2$, while,
\begin{equation}
\label{Eq:angle}
\tan\theta_{W} = \frac{g'}{g},~~~ \sin 2\alpha = \frac{4qg_x}{ \sqrt{g^{2}+{g'}^{2}} } \frac{(M_{Z}^{\text{SM}})^{2}}{M_{Z}^{2} - M_{Z'}^{2}} \,.
\end{equation}
From Eq.~\eqref{Eq:Gauge_boson_mass}, it can be seen that in the limit $u_\chi \to 0$, the $Z'$ mass $M_{Z'}$ is of order $\mathcal{O}(v_\varphi g_x)$. Therefore, the presence of at least one scalar singlet is required to allow the mass of $Z'$ to vary independently of the electroweak scale.

In the gauge-boson physical basis, the relevant part of the covariant derivative can be written as,
\begin{align}\label{Eq:codrivative2}
D_{\mu}= & \partial_{\mu} +ig \left[ T^{+} W_{\mu}^{+} + T^{-}W_{\mu}^{-} \right] + ieQA_{\mu} + i\frac{g}{\cos \theta_{W}} \left[ (T^{3} - Q\sin \theta_{W}^{2}) \cos \alpha - X\frac{\cos \theta_{W}}{g} \tilde{g}_{x}\sin \alpha   \right] Z_{\mu} \nonumber\\
& +  i \left[  \frac{g}{\cos \theta_{W}} (T^{3} - Q \sin \theta_{W}^{2}) \sin \alpha + X \tilde{g}_{x} \cos \alpha \right] Z'_{\mu}\,.
\end{align}
In this expression, $T^3$ and $Q$ denote the third component of the weak isospin and the electric charge of the fermion, respectively, while $X$ represents its $U(1)_X$ charge. 
The interaction vertices between these bosons and fermions in the fermion gauge basis
can be obtained from Eq. \eqref{Eq:Kinetic_fermion}, using the definition of the covariant derivative in Eq. \eqref{Eq:codrivative2}.
\subsection{Vector and Axial vector couplings}\label{subsec:Vector_axialV_couplings}
We now discuss the structure of vector and axial-vector couplings for SM ($A_{\mu}, W_{\mu}^{\pm},Z_{\mu}$) and BSM ($Z'_{\mu}$) vector bosons. 
As evident from Eq.~\eqref{Eq:codrivative2}, the couplings of the $W^\pm$ boson and the photon remain unchanged from their SM values.
In the neutral sector, however, gauge-boson mixing modifies the weak neutral currents.
The Lagrangian corresponding to the neutral current interactions mediated by the $Z$ and $Z'$ bosons is given by,
\begin{equation} \label{SM_lagrangian}
- \mathcal{L}_{\text{int}} \subset  \frac{g}{\cos\theta_{W}} \overline{\psi} \gamma^{\mu}\left(g_{\psi_{_{\mathtt{L}}}}^{z} P_{\mathtt{L}} + g_{\psi_{_{\mathtt{R}}}}^{z} P_{\mathtt{R}}\right)\psi Z_{\mu}+ \overline{\psi} \gamma^{\mu}\left(g_{\psi_{_{\mathtt{L}}}}^{z'} P_{\mathtt{L}}+g_{\psi_{_{\mathtt{R}}}}^{z'} P_{\mathtt{R}}\right)\psi ~Z'_{\mu} \,. 
\end{equation} 
Here, $P_{\mathtt{L(R)}}=(1 \mp \gamma^{5})/2$ are the left- and right- handed chirality projection operators.
The couplings $g_{\psi_{_{\mathtt{L/R}}}}^{z}$ and $g_{\psi_{_{\mathtt{L/R}}}}^{z'}$ in the fermion gauge basis are given by,
\begin{subequations}\label{SM_Couplings}
\begin{align}
&g_{\psi_{_{\mathtt{L}}}}^{z} = \Big(T_{\psi_{_{\mathtt{L}}}}^3 - Q_{\psi}\sin^{2}\theta_{W}\Big)\cos{\alpha} - \frac{X_{\psi_{_{\mathtt{L}}}}\tilde{g}_x}{g}\sin{\alpha}\cos{\theta_{W}}\,,\label{SM_Couplings1} \\
&g_{\psi_{_{\mathtt{R}}}}^{z} = - Q_{\psi}\sin^{2}\theta_{W}\cos{\alpha} - \frac{X_{\psi_{_{\mathtt{R}}}}\tilde{g}_x}{g}\sin{\alpha}\cos{\theta_{W}}\, ,\label{SM_Couplings2}\\
& g_{\psi_{_{\mathtt{L}}}}^{z'} =  \frac{g}{\cos \theta_{W}} \Big(T_{\psi_{_{\mathtt{L}}}}^3 - Q_{\psi} \sin^{2} \theta_{W} \Big)\sin{\alpha} + X_{\psi_{_{\mathtt{L}}}}\tilde{g}_x\cos{\alpha}\,,\label{SM_Couplings3} \\
& g_{\psi_{_{\mathtt{R}}}}^{z'} =   \frac{g}{\cos \theta_{W}} \Big( - Q_{\psi} \sin^{2} \theta_{W} \Big)\sin{\alpha} + X_{\psi_{_{\mathtt{R}}}}\tilde{g}_x\cos{\alpha}\,\label{SM_Couplings4}.
\end{align}
\end{subequations}
Eqs.~\eqref{SM_Couplings1} and \eqref{SM_Couplings2} represent the 
physical $Z$ couplings with left- and right-handed SM fermions, 
respectively, while Eqs.~\eqref{SM_Couplings3} and \eqref{SM_Couplings4} 
correspond to the $Z'$ ones. $T^3_{\psi_{_{\mathtt{L}}}}$ and $Q_{\psi}$ 
denote the third component of the weak isospin and the electric charge 
of the fermion $\psi$, while $X_{\psi_{_{\mathtt{L/R}}}}$ are the respective $U(1)_X$ 
charges.
Notably, the $Z$ boson couplings to SM fermions acquire BSM contributions due to gauge-boson mixing.  
These additional terms are proportional to the effective $U(1)_X$ gauge coupling $\tilde{g_x}$ and to $\sin\alpha$.
The mixing angle $\alpha$ could be constrained by the precisely measured electroweak parameter $\rho = M_{W}^{2}/(M_{Z}^{2}\cos^{2}\theta_{W})$ \cite{Ross:1975fq,Bento:2023flt,Bento:2023weq}, which equals 1 at tree level in the SM. The deviations from unity in ${\rm SM} \otimes {\rm U}(1)_{X}$ theories can be expressed as~\cite{Bento:2023flt},
\begin{equation}
\label{Eq:Rho_parameter}
    \rho -1 = \left[  \left( \frac{M_{Z'}}{M_{Z}} \right)^{2} -1    \right] \sin^{2} \alpha\,.
\end{equation}
Using the $3\sigma$ allowed range of the $\rho$ parameter, $\rho = 1.00038 \pm 0.00060$ \cite{ParticleDataGroup:2020ssz}, obtained from the most recent global fit to electroweak precision data, implies $q g_x \lesssim 5.5 \times 10^{-3}$ for sufficiently light $Z'$ ($M_{Z'}^{2}/M_{Z}^{2} \ll 1$) \cite{Majumdar:2024dms,Prajapati:2026tfv}.
Under this limit, Eqs.~\eqref{Eq:VEV_Paramter_ux}, 
\eqref{Eq:angle}, and \eqref{Eq:Rho_parameter} 
can be approximated as,
\begin{equation}\label{Eq:approx_light_zp}
    u_{\chi} \simeq \frac{M_{Z'}}{g_x}\,,~~~ \alpha \simeq \frac{2 \cos{\theta_{W}}  }{g} q g_x\,,~~~ \rho \simeq 1 - \alpha^2\,.
 \end{equation}
Notably, $\alpha$ becomes independent of $M_{Z'}$ in this limit. Using the above approximations, the couplings in Eq. \eqref{SM_Couplings} can be simplified. By taking $\cos\alpha \simeq 1$, the BSM contribution to the $Z$ boson couplings becomes proportional to $q\, g_x \tilde{g}_x$, which can be safely neglected with respect to the SM contribution. Consequently, $Z$ couplings remain identical to those in the SM. In this limit, the couplings of fermions to the $Z'$ boson are expressed as,
\begin{subequations}\label{EQ:Zp Couplings}
\begin{align}
&g_{\psi_{_{\mathtt{L}}}}^{z'} \simeq  \left[2 X_{\Phi} \Big(T_{\psi_{_{\mathtt{L}}}}^3 - Q_{\psi} \sin^{2} \theta_{W} \Big)+ X_{\psi_{_{\mathtt{L}}}} \right]g_x + 2Q_{\psi}\tilde{g} \cos^{2} \theta_{W}\,,\, \\
&  g_{\psi_{_{\mathtt{R}}}}^{z'} \simeq \left[   2 X_{\Phi}\Big( - Q_{\psi} \sin^{2} \theta_{W} \Big) + X_{\psi_{_{\mathtt{R}}}} \right] g_x + 2Q_{\psi}\tilde{g} \cos^{2} \theta_{W}\,.
\end{align}
\end{subequations}
From Eq.~\eqref{SM_lagrangian}, the vector and axial vector couplings of the fermion $\psi$ with the $Z'$ boson in the fermion gauge basis are defined as,
\begin{equation}\label{Eq:Vector_axialVector_Coupling}
    C^{\psi}_{V} = \frac{g_{\psi_{_{\mathtt{R}}}}^{z'} + g_{\psi_{_{\mathtt{L}}}}^{z'}}{2},~~~~ C^{\psi}_{A} =  \frac{g_{\psi_{_{\mathtt{R}}}}^{z'} - g_{\psi_{_{\mathtt{L}}}}^{z'}}{2}\,,
\end{equation}
which, using Eq.~\eqref{EQ:Zp Couplings}, reduce to
\begin{equation}\label{Eq:VA_coup}
C^{\psi}_{V}  \simeq \left[  X_{\Phi} (T_{\psi_{_{\mathtt{L}}}}^{3} - 2Q_{\psi} \sin^{2}{\theta_{W}}) + C^{\psi}_{V(0)}\right]g_x+ 2Q_{\psi}\tilde{g} \cos^{2} \theta_{W},~~~~ 
C^{\psi}_{A} \simeq \left[ - X_{\Phi} T_{\psi_{_{\mathtt{L}}}}^{3} + C^{\psi}_{A(0)} \right]g_x\,. 
\end{equation}
Here, $C^{\psi}_{V(0)/A(0)} = (X_{\psi_{\mathtt{R}}} \pm X_{\psi_{\mathtt{L}}})/2$ denote the vector and axial-vector charges of the $Z'$ to the fermion $\psi$ in the limit where the Higgs $U(1)_X$ charge and kinetic mixing is zero.

At this point, it is worth emphasizing some key properties of these couplings:
\begin{itemize}
\item The Higgs doublet $\Phi$, having the larger VEV, effectively induces $Z-Z'$ mixing. Consequently, a factor proportional to its charge $X_{\Phi}$ appears in Eq.~\eqref{Eq:VA_coup}. However, its charge $X_\Phi$ is not a free parameter, since it is constrained by the mass generation mechanism of the SM fermions. The Yukawa sector responsible for generating SM fermion masses can be expressed as,
\begin{equation}\label{Eq:SM_Yukawa_couplings}
   - \mathcal{L_{Y}} = Y_{e}^{ij}\overline{L^{i}} H e_{_{\mathtt{R}}}^{j} +Y_{u}^{ij} \overline{Q^{i}} \tilde{H} u_{_{\mathtt{R}}}^{j}  + Y_{d}^{ij} \overline{Q^{i}} H d_{_{\mathtt{R}}}^{j} + \text{h.c.}\,,  
\end{equation}
where, in a two Higgs doublet scenario, $H$ represents $\Phi$ or $\varphi$. Requiring the corresponding Yukawa couplings to be gauge invariant imposes the following constraints on the $U(1)_X$ charges of $H$,
\begin{equation}\label{Eq:Higgs_Charge_Constraints}
    X_{H} = \frac{X_{\psi_{\mathtt{R}}}^{j} - X_{\psi_{\mathtt{L}}}^{i}}{2 T^{3}_{\psi_{_{\mathtt{L}}}}}\,.
\end{equation}
\item It is clear from Eq.~\eqref{Eq:Higgs_Charge_Constraints} that, if the Higgs doublet $\Phi$, which acquires the larger VEV, also generates the mass of a fermion $\psi$ through Yukawa interactions, then its $U(1)_X$ charge is $X_{\Phi}=C^{\psi}_{A(0)}/T^{3}_{\psi_{_{\mathtt{L}}}}$. Inserting this into Eq.~\eqref{Eq:VA_coup}, one finds that the corresponding axial-vector coupling of the $Z'$ to the fermion $\psi$ 
vanishes, rendering the interaction purely vector. Therefore, within this setup, to obtain a nonzero $Z'$ axial-vector coupling to SM fermion $\psi$, its mass must be generated by the Higgs doublet with the smaller VEV ($\varphi$). This condition demands generation-specific $U(1)_X$ charge assignments for SM fermions, ensuring that only $\varphi$ couples to the corresponding fermion through Yukawa interactions.
\item As discussed in Sec. \ref{Sec:Exp_Constraints}, for light $Z'$ ($ M_{Z'}\sim 17$ MeV), strong constraints comes from neutrino-electron and neutrino-nucleus scattering. These can be alleviated if the $Z'$ couplings to electrons or neutrinos are sufficiently suppressed or vanish. From Eq.~\eqref{EQ:Zp Couplings}, one can see that in the present framework the neutrino-$Z'$ vanishes for $X_{\Phi} = -X_{L}$.
\end{itemize}
It should be noted that the couplings discussed above are expressed in the fermion weak basis. In the case of generation specific SM fermion charges, the rotation to the mass basis may modify these couplings or/and even generate FCNCs. Nonetheless, appropriate choices of the fermionic $U(1)_X$ charges and rotation matrices 
can be made such that the couplings remain identical to those given in Eq.~\eqref{Eq:VA_coup} -- see Sec. \ref{Sec:Model}. Consequently, in a two Higgs doublet framework where first-generation SM fermions couple exclusively to the Higgs doublet with the smaller VEV, one can generate the axial-vector $Z'$ couplings required to explain the ATOMKI anomaly. 
\subsection{$Z'$ couplings with the neutron and proton}

Having discussed the $Z'$ couplings to elementary fermions, we now focus on the definition of the effective $Z'$ couplings with neutrons and protons. The vector and axial-vector coupling of the nucleon $N$ with $Z'$ ($C_N$ and $a_N$, respectively) can be written as \cite{Barducci:2022lqd},
\begin{equation}
    -\mathcal{L} = C_N\,
 \overline{N}\gamma^{\mu}N Z'_{\mu} + a_N \, \overline{N}\gamma^{\mu}\gamma^{5}NZ'_{\mu}\,. \end{equation}
For the proton ($N\equiv p$) and neutron ($N\equiv n$) the vector couplings are~\cite{Barducci:2022lqd},
\begin{equation}\label{Eq:Proton_Neutron_Vector_coupnig_def}
    C_{p} = 2C_{V}^{u} + C_{V}^{d}\,,~ C_{n} = C_{V}^{u} + 2C_{V}^{d}\,,
\end{equation}
while the axial-vector ones are given by the sum over quark axial couplings $C_A^q$, weighted by the fraction of the nucleon spin carried by each quark $\Delta q^{(N)}$, namely~\cite{Kozaczuk:2016nma,Bishara:2016hek},
\begin{equation}\label{Eq:Proton_Neutron_Axial_coupnig_def}
 a_N = \sum_{q} \Delta_{q}^{(N)}C_{A}^{q}\;,\; N\equiv n,p\,,
\end{equation}
with spin-fraction values~\cite{Kozaczuk:2016nma,Bishara:2016hek,Barducci:2022lqd},
\begin{equation}
    \Delta_{u}^{p} = \Delta_{d}^{n} = 0.897(27),~  \Delta_{d}^{p} = \Delta_{u}^{n} = -0.367(27),~\Delta_{s}^{p} = \Delta_{s}^{n} = -0.026(4)\,.
\end{equation}
As for the contributions from heavy quarks, these are small and can be safely neglected.

To explain the ATOMKI anomaly, the axial-vector couplings of the proton and neutron are required to be $a_{n,p} \approx 10^{-4} - 10^{-3}$. Furthermore, these couplings must remain compatible with several experimental constraints, including atomic parity violation, beam-dump experiments, anomalous magnetic moments $(g-2)$, meson decays, neutrino-electron scattering, and LHC collider searches. In the following section, we provide a detailed discussion of these constraints and their implications for the $U(1)_X$ parameter space.  

\section{Experimental constraints on a light spin-1 boson}\label{Sec:Exp_Constraints}
For our $U(1)_X$ gauge extension of the SM, the interaction between $Z'$ and a fermion $\psi$ given in Eq.~\eqref{SM_lagrangian} can be rewritten as,
\begin{equation}\label{IntZP2}
-\mathcal{L}_{\text{int}}^{\psi}= 
\overline{\psi}\gamma^{\mu}(C_{V}^{\psi} +\gamma
^{5}C_{A}^{\psi})\psi Z'_{\mu}\;\;,\;\; C^{\psi}_{V/A} = (g_{\psi_{_{\mathtt{R}}}}^{z'} \pm g_{\psi_{_{\mathtt{L}}}}^{z'})/2\,,
\end{equation}
where we assume a diagonal coupling in the mass basis of fermions. From the fundamental fermion couplings to the $Z'$ boson, the effective vector ($C_{p,n}$) and axial-vector ($a_{p,n}$) nucleon couplings can be computed using Eqs.~\eqref{Eq:Proton_Neutron_Vector_coupnig_def} and \eqref{Eq:Proton_Neutron_Axial_coupnig_def}. Many of the experimental constraints on the various couplings involved have been previously discussed in the literature~\cite{Feng:2016jff,Feng:2016ysn,Barducci:2022lqd}. For completeness, we summarize them here.
\begin{enumerate}
  \item \textbf{Prompt decay in the ATOMKI detector :} To produce the IPC signal, the $Z'$ boson must decay within the ATOMKI detector. The requirement of a prompt decay inside the detector imposes the following constraints,
\begin{equation}
\sqrt{(C_{V}^{e})^{2} + (C^{e}_{A})^{2}} \gtrsim 3\times10^{-7} \times \sqrt{\text{Br}(Z' \to e^{+}e^{-})}\,.
\end{equation}
\item \textbf{Atomic parity violation :} A light $Z'$ that couples to both electrons and nucleons can contribute to parity violation in atomic transitions. The effective low-energy operators mediating such processes can be written as \cite{Barducci:2022lqd},
\begin{equation}
-\mathcal{L} = \frac{1}{M_{Z'}^2} \left[ C_V^u C_A^e \, (\bar{u} \gamma^\mu u) (\bar{e} \gamma_\mu \gamma^5 e) + C_A^u C_V^e \, (\bar{u} \gamma^\mu \gamma^5 u) (\bar{e} \gamma_\mu e) + (u \leftrightarrow d) \right],
\end{equation}
where the last term denotes the analogous contributions from down quarks.
Among these effective interactions, only the the terms which couple axial-vector electron current to the vector quark currents add coherently and contribute to the total weak charge of the nucleus~\cite{Ginges:2003qt,Bouchiat:2004sp}. Measurements of the effective weak charge in $^{133}$Cs~\cite{Porsev:2009pr}, combined with precise SM predictions, yield the following stringent bounds~\cite{Dzuba:2012kx,Arcadi:2019uif,Barducci:2022lqd},
\begin{equation}\label{Eq:Cons_APV}
  \lvert C_{A}^{e}\rvert \left\lvert  \frac{188}{399} C_{V}^{u} + \frac{211}{399} C_{V}^{d}  \right\rvert \lesssim 1.8 \times 10^{-12}\,.
\end{equation}
\item \textbf{Parity violation in M\o ller Scattering :} Precision measurements of parity violation in fixed-target electron-electron (Møller) scattering from the SLAC E158 experiment provide one of the most sensitive probes at low momentum transfer, $Q^2 = 0.0026~\text{GeV}^2$~\cite{SLACE158:2005uay}. 
Recasting these bounds on the product of the electron vector and axial-vector couplings, we get~\cite{Kahn:2016vjr},
\begin{equation}\label{Eq:Cons_APV_Moller}
    \lvert C_{V}^{e}  C_{A}^{e} \rvert \lesssim 10^{-8}\,.
\end{equation}
\item \textbf{Magnetic Moment of electron and muon :} Light vector or axial vector mediator coupling to electron/muon gives a contribution to $a_{e/\mu}= (g-2)_{e/\mu}/2$.
The theoretically predicted value of $a_e$ using the 
measurements of the fine-structure constant from $^{87}\text{Rb}$ \cite{gm2Rb} and $^{137}\text{Cs}$ \cite{doi:10.1126/science.aap7706} is $1.6 \sigma$ lower and $-2.4 \sigma$ higher, respectively than the direct experimental measurement of $a_e$ \cite{PhysRevLett.100.120801},
\begin{equation}
    \begin{split}
       & \Delta a_{e} (\text{Rb}) = a_{e}^{\text{exp}}- a_{e}^{\text{th}} = (4.8 \pm 3.0)\times 10^{-13}\,, \\
       & \Delta a_{e} (\text{Cs}) = a_{e}^{\text{exp}}- a_{e}^{\text{th}} = (-8.8 \pm 3.6 )\times 10^{-13}\,.
    \end{split}
\end{equation}
Similarly, following the 2025 white paper~\cite{Aliberti:2025beg}, which updated the theoretical SM prediction, the deviation between the experimental world average and the SM value is,
\begin{equation}
    \Delta a_{\mu} = a_{\mu}^{\text{exp}}-a_{\mu}^{\text{WP2025}} = 3.8 (6.3) \times 10^{-10}\,.
\end{equation}
Imposing that the BSM contribution from the $Z'$ remains within the 
experimental uncertainty yields the following constraints \cite{Jegerlehner:2009ry},
\begin{subequations}\label{Eq:cons:gm2}
\begin{align}
& \Delta a_{e}^{\text{BSM}} (\text{Rb}) \simeq 7.63  \times 10^{-6} (C_{V}^{e})^{2} -3.85\times 10^{-5} (C_{A}^{e})^{2} \in \left[ 0 - 7.8\times 10^{-13} \right]\,,\\
& \Delta a_{e}^{\text{BSM}} (\text{Cs}) \simeq 7.63 \times 10^{-6}(C_{V}^{e})^{2} -3.85\times 10^{-5} (C_{A}^{e})^{2}\in \left[ -12.4\times 10^{-13} - 0 \right]\,,\\
& \Delta a_{\mu}^{\text{BSM}}  \simeq 2.51 \times 10^{-4} (C_{V}^{\mu})^{2}-0.66\times (C_{A}^{\mu})^{2}\in \left[ 0-10.1\times 10^{-10} \right]\,.
\end{align}
\end{subequations}
\item \textbf{Electron positron annihilation into $Z'$ and a photon ($\mathbf{e^{+}e^{-} \to Z' \gamma}$) :} The KLOE experiment searches for the process $e^+ e^- \to Z' \gamma$ followed by $Z' \to e^+ e^-$~\cite{Anastasi:2015qla}. Neglecting kinematic differences with respect to the pure dark photon case, Ref.~\cite{Barducci:2022lqd} recasts these limits as,
\begin{equation}\label{Eq:Const_epen_scattering}    \sqrt{(C_{V}^{e})^{2}+(C_{A}^{e})^{2}}\lesssim \frac{6.1 \times 10^{-4}}{\sqrt{\text{BR}(Z' \to e^{+}e^{-})}}\,.
\end{equation}
\item \textbf{Beam dump experiments :} Electron beam dump experiments, such as SLAC E141 \cite{PhysRevLett.59.755}, SLAC E137 \cite{Bjorken:1988as}, Fermilab E774 \cite{Bross:1989mp}, KEK \cite{PhysRevLett.57.659} , and NA64 \cite{NA64:2018lsq,NA64:2019auh,MolinaBueno:2021dqu}, search for $Z'$ production via bremsstrahlung from electrons scattering off target nuclei \cite{Bjorken:2009mm,Essig:2013lka,Fabbrichesi:2020wbt}. These experiments constrain the electron coupling with $Z'$ in two complementary regimes: either the coupling is sufficiently weak to avoid significant $Z'$ production, or it is sufficiently strong that the $Z'$ decays inside the dump, with its decay products absorbed before reaching the detector. This leads to the following two limits \cite{Feng:2016ysn,Barducci:2022lqd},
\begin{equation}\label{Eq:Cons_Beam_dump}
\sqrt{(C_{V}^{e})^{2}  + (C_{A}^{e})^{2}} \lesssim 1.1 \times 10^{-8}\;\text{or}\; \sqrt{(C_{V}^{e})^{2}  + (C_{A}^{e})^{2}} \gtrsim 3.6 \times 10^{-5} \times \sqrt{\text{BR}(Z' \to e^{+}e^{-})}\,.
\end{equation}
\item \textbf{Pion decay constraints (SINDRUM-I) :} The SINDRUM-I experiment employed a spectrometer originally designed to search for the lepton-number-violating process $\mu^+ \to e^+ e^+ e^-$~\cite{10.21468/SciPostPhysProc.5.007}. It also studied a variety of other processes with high precision, including the most accurate measurement to date of the rare charged pion decay $\pi^+ \to e^+ \nu_e \, e^+ e^-$~\cite{SINDRUM:1989qan}. 
For a light scalar BSM particle in the mass range $10~\text{MeV} < M_X < 110~\text{MeV}$, the SINDRUM-I experiment constrains the branching ratio $\text{BR}(X \to e^+ e^-)$ to below $\mathcal{O}(10^{-9})-\mathcal{O}(10^{-11})$~\cite{SINDRUM:1989qan}.
Since such a light scalar exhibits kinematics similar to the longitudinal mode of a light $Z'$ boson, Ref.~\cite{Hostert:2023tkg} recasts these limits as constraints on the net effective charge with which the $Z'$ couples,
\begin{equation}\label{Eq:Const_SINDRUM}
    \lvert \Delta Q_{Z'} \rvert \lesssim \frac{8.5 \times 10^{-5}}{\sqrt{\text{BR}(Z' \to e^{+} e^{-})}}\,~~\text{at}\,\, 90 \%\,\,\text{C.L.}\,,
\end{equation}
with $\Delta Q_{Z'} = (C_{V}^{u}+C_{A}^{u})-(C_{V}^{d}+C_{A}^{d}) - (C_{V}^{\nu}-C_{A}^{\nu}) + (C_{V}^{e}-C_{A}^{e})$.
\item \textbf{Protophobic nature of vector coupling $\mathbf{(\pi^{0} \to \gamma Z')}$ :}  In the purely vector case, the strongest bounds arise from the NA48 experiment's searches for $\pi^{0} \to \gamma (Z' \to e^{+} e^{-})$ decays, in dark photon searches \cite{NA482:2015wmo}. 
These constraints require the $Z'$ to exhibit protophobic vector couplings to quarks \cite{Feng:2016jff,Feng:2016ysn,Barducci:2022lqd},
\begin{equation}\label{Eq:Const_Protophobic}
    \lvert C_{p} \rvert \times \sqrt{\text{BR}(Z' \to e^{+} e^{-})} \lesssim 2.5 \times 10^{-4}\,.
\end{equation}
\item \textbf{Neutron lead scattering :}  The angular distribution in neutron lead scattering provides bounds on new light mediators, expressed as \cite{Barbieri:1975xy, Feng:2016ysn,Barducci:2022lqd},
\begin{equation}\label{Eq:Const_Neutron_lead}
    \lvert C_{n} \rvert \left\lvert \frac{126}{208}C_{n} + \frac{82}{208} C_{p} \right\rvert \lesssim 3.6 \times 10^{-5}\,.
\end{equation}
\item \textbf{Neutrino electron (nucleus) scattering :} One of the most stringent limits comes from the
neutrino-electron ($\nu-e$) elastic scattering and coherent elastic neutrino-nucleus scattering (CE$\nu$NS)~\cite{AristizabalSierra:2018eqm, Majumdar:2021vdw, Coloma:2022avw, Coloma:2022umy, A:2022acy, Majumdar:2024dms, Blanco-Mas:2024ale, DeRomeri:2024iaw, Chattaraj:2025rtj, Rathsman:2026smv}.
These experiments constrain the product of the neutrino and target fermion couplings, typically expressed as $\sqrt{|C_{V,A}^{\nu} C_{V,A}^{e}|}$ for electron scattering and $\sqrt{|C_{V,A}^{\nu} C_{V,A}^{q}|}$ for nuclear scattering. Given the $\sim 17$~MeV mass of the $Z'$, these constraints are particularly robust, typically yielding upper bounds in the range of $ \lesssim \mathcal{O}(10^{-5}) - \mathcal{O}(10^{-4})$.
\end{enumerate}

As previously noted, addressing the ATOMKI anomalies requires axial-vector couplings to the proton and neutron of the order $a_{n,p} \sim 10^{-4}-10^{-3}$. 
Satisfying this requirement while simultaneously evading the stringent experimental constraints discussed above presents a nontrivial challenge for any UV complete framework. In the following, we present a 
fully consistent, anomaly-free model that successfully accommodates the ATOMKI signals while remaining compatible with all existing experimental bounds.
%

\section{$U(1)_X$ model for ATOMKI anomalies}\label{Sec:Model}
In this section, we construct an explicit model that realizes the fermion coupling structure derived in Sec.~\ref{subsec:Vector_axialV_couplings}, while remaining consistent with the stringent constraints on a light $Z'$ boson as detailed in Sec.~\ref{Sec:Exp_Constraints}. For this purpose, we first need to fix the $U(1)_X$ charges of fermions and scalars. As noted earlier, the $U(1)_X$ charges of the fermions are not arbitrary; they must satisfy gauge anomaly cancellation conditions to ensure the theory remains unitary and renormalizable~\cite{Adler:1969gk,Bardeen:1969md,Bell:1969ts,Delbourgo:1972xb,Alvarez-Gaume:1983ihn,Witten:1982fp}. Apart from the anomaly-cancellation conditions related to the SM gauge groups, 
the extended gauge symmetry $\text{SM} \otimes U(1)_X$ generates additional 
triangle anomalies associated with the $U(1)_X$ current. 
The corresponding conditions are summarized below :
\begin{subequations}\label{Eq:Anomaly_can_con}
\begin{align}
&[SU(3)_{C}]^2[U(1)_X]= \sum_{i=1}^{3}\left(2 X_{Q^{^{i}}} -  X_{u_{_{\mathtt{R}}}^{^{i}}}-X_{d_{_{\mathtt{R}}}^{^{i}}}\right) \label{Eq:anoUx1} = 0\,, 
\\
&[SU(2)_{\mathtt{L}}]^2[U(1)_X]= \sum_{i=1}^{3}\left( X_{L^{^{i}}} +3 X_{Q^{^{i}}}\right) = 0\label{Eq:anoUx2}\,,
\\
&[U(1)_{Y}]^2 [U(1)_X]= \sum_{i=1}^{3} \left( X_{L^{^{i}}} + \frac{1}{3}  X_{Q^{^{i}}} -2  X_{e_{_{\mathtt{R}}}^{^{i}}} -\frac{8}{3}X_{u_{_{\mathtt{R}}}^{^{i}}}-\frac{2}{3}  X_{d_{_{\mathtt{R}}}^{^{i}}} \right) = 0\label{Eq:anoUx3}\,, 
\\
&[U(1)_{Y}] [U(1)_X]^2=  \sum_{i=1}^{3} \left\{   \left(X_{Q^{^{i}}}\right)^{2}-\left(X_{L^{^{i}}}\right)^{2}  + \left(X_{e_{_{\mathtt{R}}}^{^{i}}}\right)^2 -2 \left(X_{u_{_{\mathtt{R}}}^{^{i}}}\right)^2 + \left(X_{d_{_{\mathtt{R}}}^{^{i}}}\right)^2  \right\} = 0 \label{Eq:anoUx4} \,,
\\ 
& [U(1)_X]^3= \sum_{i=1}^{3} \left[ 2\left(X_{L^{^{i}}}\right)^{3}  +6 \left(X_{Q^{^{i}}}\right)^{3}   - \left(X_{e_{_{\mathtt{R}}}^{^{i}}}\right)^{3} -3 \left\{ \left(X_{u_{_{\mathtt{R}}}^{^{i}}}\right)^{3}  + \left(X_{d_{_{\mathtt{R}}}^{^{i}}}\right)^{3} \right\} \right]  - \sum_{i=1}^{3} \left(X_{\nu_{_{\mathtt{R}}}^{i}}\right)^{3} = 0 \label{Eq:anoUx5} \,,
\\&[G]^2[U(1)_X]= \sum_{i=1}^{3} \left\{ 2X_{L^{^{i}}}  + 6 X_{Q^{^{i}}} -X_{e_{_{\mathtt{R}}}^{^{i}}}  -3\left( X_{u_{_{\mathtt{R}}}^{^{i}}} + X_{d_{_{\mathtt{R}}}^{^{i}}}\right) \right\}  - \sum_{i=1}^{3} X_{\nu_{_{\mathtt{R}}}^{i}} = 0 \label{Eq:anoUx6}\,.
\end{align}
\end{subequations}
In the above expressions, $X_{\psi}$ corresponds to the  $U(1)_X$ charge of the fermion $\psi$. 
The generation indices are denoted by $i,j = 1,2,3$. 
The $SU(2)_L$ lepton and quark doublets are given by 
$L = (\nu_{\mathtt{L}}, e_{\mathtt{L}})^{T}$ and 
$Q = (u_{\mathtt{L}}, d_{\mathtt{L}})^{T}$, 
whereas the corresponding $SU(2)_L$ singlets are 
$l = e_{\mathtt{R}}$ and 
$q = (u_{\mathtt{R}}, d_{\mathtt{R}})$.
For the consistency of the theory, all six conditions in Eq.~\eqref{Eq:Anomaly_can_con},
along with the anomaly cancellation conditions involving purely SM currents, 
must be simultaneously satisfied.
To satisfy these requirements, we introduce three SM 
right-handed neutrino singlets, $\nu_{\mathtt{R}}^{i}$ ($i=1,2,3$), 
each assigned a $U(1)_X$ charge $X_{\nu_{\mathtt{R}}^{i}}$.

Within this setup, to realize both vector and axial-vector couplings 
of the $Z'$ to SM fermions, we introduce generation specific 
$U(1)_X$ charge assignments, as discussed in Sec.~\ref{subsec:Vector_axialV_couplings}.
Furthermore, we solve the anomaly cancellation conditions while taking into account 
the constraints on a light $Z'$ boson with mass $\approx 17\,\text{MeV}$. 
These experimental constraints are discussed in Sec.~\ref{Sec:Exp_Constraints}. 
Very stringent bounds on the electron-$Z'$ coupling arise from atomic parity violation, and M\o ller scattering ~\cite{Dzuba:2012kx,Kahn:2016vjr,Arcadi:2019uif,Barducci:2022lqd}. To evade these bounds, we impose 
$C_{A}^{e} = 0$, implying that the electron must couple to the Higgs doublet with the larger VEV through the Yukawa interaction.
Furthermore, to avoid constraints from neutrino electron or neutrino nucleus scattering \cite{Majumdar:2021vdw,Coloma:2022avw,A:2022acy, Majumdar:2024dms,Blanco-Mas:2024ale,DeRomeri:2024iaw,Chattaraj:2025rtj}, we assign opposite $U(1)_X$ charges to the lepton doublet and the 
Higgs doublet with the larger VEV, i.e., $X_{L_i} = -X_{\Phi} \equiv X_{L}$. As a result, the neutrino $Z'$ coupling vanishes, see the discussion after eq. \eqref{Eq:Higgs_Charge_Constraints} in Sec. \ref{subsec:Vector_axialV_couplings}.
In addition, to evade the stringent bounds from charged pion decays 
$\pi^+ \to e^+ \nu_e \, e^+ e^-$ reported by the SINDRUM-I experiment 
\cite{SINDRUM:1989qan,Hostert:2023tkg}, we impose $\Delta Q_{Z'} = 0$, see Sec.~\ref{Sec:Exp_Constraints} for details.
The resulting $U(1)_X$ charge assignments are summarized in 
Table~\ref{Tab:parametertable1}, where $X_{L}$ and $X_{\nu}$ are free parameters.
\begin{table}[ht]
\centering
\renewcommand{\arraystretch}{2}
\setlength{\tabcolsep}{20pt} 
\small
\begin{tabular}{|c c|c c|}
\hline
Fields & $U(1)_X$ & Fields & $U(1)_X$ \\ 
\hline
$Q^{1},Q^{2},Q^{3}$ & $-\frac{X_{L}}{3}$ & $L^{1},L^{2},L^{3}$ & $X_{L}$ \\ 
$u_{\mathtt{R}}^{1},u_{\mathtt{R}}^{2},u_{\mathtt{R}}^{3}$ & $\left(-\frac{4X_{L}}{3},~ X_{\nu} - \frac{4X_{L}}{3} ,~-\frac{4X_{L}}{3} \right)$ & $e_{\mathtt{R}}^{1},e_{\mathtt{R}}^{2},e_{\mathtt{R}}^{3}$ & $(2X_{L},~2X_{L},~2X_{L}-X_{\nu})$ \\ 
$d_{\mathtt{R}}^{1},d_{\mathtt{R}}^{2},d_{\mathtt{R}}^{3}$ & $\left(\frac{2X_{L}}{3},~-X_{\nu} + \frac{2X_{L}}{3},~\frac{2X_{L}}{3} \right)$ & $\nu_{\mathtt{R}}^{1},\nu_{\mathtt{R}}^{2},\nu_{\mathtt{R}}^{3}$ & $(0,~0,~X_{\nu})$ \\ 
\hline
$\Phi$ & $-X_{L}$ & $\chi$ & $ -X_{\nu} $ \\
$\varphi$ & $X_{\nu} - X_{L}$ & & \\
\hline
\end{tabular}
\caption{Charge assignments of SM and BSM fields under the $U(1)_X$ symmetry. Both scalar $\Phi$ and $\varphi$ participate in generating the masses for SM fermions via Yukawa interactions. The singlet $\chi$ allows the $Z'$ mass to vary freely, and its $U(1)_X$ charge is chosen such that no massless Goldstone bosons appear in the model. $X_L$ and $X_\nu$ are free parameters.}
\label{Tab:parametertable1}
\end{table}

The $U(1)_X$ charges of the lepton and quark $SU(2)_L$ doublets are flavor universal. In contrast, the quark and lepton singlets 
carry generation specific $U(1)_X$ charges, with two generations 
sharing identical charges and the remaining generation assigned 
a distinct $U(1)_X$ charge.
Since the $U(1)_X$ assignments for the SM fermions are chiral, the Higgs sector must also be charged under $U(1)_X$ to allow for gauge-invariant Yukawa interactions that generate fermion masses. 
In this model, we employ two Higgs doublets, $\Phi$ and $\varphi$, with 
$U(1)_X$ charges $-X_L$ and $X_\nu - X_L$, respectively. Additionally, a complex scalar singlet $\chi$ is introduced to ensure that the $Z'$ mass remains independent of the electroweak scale, see the discussion in Sec.~\ref{Sec:axialcoupling} after Eq. \eqref{Eq:angle}.
The scalar singlet $\chi$ also participates in neutrino mass generation and mixing.
In the following subsection, we provide a detailed discussion of the specific mechanisms for fermion mass generation and mixing.
\subsection{Fermion masses and mixing}
The scalar sector consists of two Higgs doublets, $\Phi$ and $\varphi$, 
and one SM singlet scalar, $\chi$.
Both scalar doublets participate in the generation of SM fermion 
masses through Yukawa interactions.
The corresponding Lagrangian density responsible for generating 
the charged fermion masses can be written as,
\begin{align}
   - \mathcal{L_{Y}} \supset & ~ Y_{e}^{i1(2)}\overline{L^{i}} \Phi e_{_{\mathtt{R}}}^{1(2)} + Y_{e}^{i3}\overline{L^{i}} \varphi e_{_{\mathtt{R}}}^{3} +Y_{u}^{i1(3)} \overline{Q^{i}} \tilde{\Phi} u_{_{\mathtt{R}}}^{1(3)}+Y_{u}^{i2} \overline{Q^{i}} \tilde{\varphi} u_{_{\mathtt{R}}}^{2} + Y_{d}^{i2} \overline{Q^{i}} \varphi d_{_{\mathtt{R}}}^{2}\nonumber \\ 
   & + Y_{d}^{i1(3)} \overline{Q^{i}} \Phi d_{_{\mathtt{R}}}^{1(3)} + \text{h.c.}\,.
\end{align}
Here, $\tilde{\Phi} (\tilde{\varphi}) = i \sigma_{2} \Phi^{*} (\varphi^{*})$, where $\sigma_{2}$ is the second Pauli matrix, and $i= 1,2,3$. The neutral components of both scalar doublets acquire VEVs as defined in Eq.~\eqref{VEV}. After $U(1)_X$ and electroweak symmetry breaking, Dirac mass matrices are generated.
Fermion masses are obtained after a biunitary diagonalization of those matrices through the field rotations: $ \psi_{_\mathtt{L}}^{m} = U_{\psi_{_\mathtt{L}}}^{\dagger}\psi_{_\mathtt{L}},\,\,  \psi_{_\mathtt{R}}^{m} = V_{\psi_{_\mathtt{R}}}^{\dagger}\psi_{_\mathtt{R}}.$ We consider the charged-lepton mass matrix to be diagonal, thereby avoiding $Z'$ mediated FCNCs in the charged-lepton sector. For quarks, there is enough freedom to generate CKM matrix of charged current interaction, defined as, 
\begin{equation}
    -\mathscr{L}_{WCC}^{q}= \frac{g}{\sqrt{2}} \left[  \overline{d_{_\mathtt{L}}^{m}} \gamma^{\mu} V_{CKM}^{\dagger} u_{_\mathtt{L}}^{m}W_{\mu}^{-} + \overline{u_{_\mathtt{L}}^{m}} \gamma^{\mu} V_{\rm CKM} d_{_\mathtt{L}}^{m}W_{\mu}^{+}  \right].
\end{equation}
Here, $d_{_\mathtt{L}}^{m} = ( d, s, b)^{T} $,  $u_{_\mathtt{L}}^{m} = (u,c,t)^{T}$, and $V_{\rm CKM} =U_{u_{_\mathtt{L}}}^{\dagger}U_{d_{_\mathtt{L}}} $. We work in standard parametrization where we take the basis where the up quark mass matrix is diagonal, $U_{u_{_\mathtt{L}}} =V_{u_{_\mathtt{R}}} = \mathtt{1}$. 
Furthermore, the flavor-universal $U(1)_X$ charges of the quark doublets ensure that no FCNCs arise from the left-handed down quarks. To prevent FCNCs from the right-handed down type quarks, we set $V_{d_R} = \mathtt{1}$. Hence, no $Z'$ mediated FCNCs arise in the charged lepton or quark sectors. Consequently, the $Z'$ couplings to these fermions remain unchanged from those given in Eq.~\eqref{Eq:VA_coup}, even in the mass basis.

Coming to neutrino masses, the relevant Lagrangian terms for neutrino mass generation are:
\begin{equation}
\begin{split}
- \mathcal{L_{Y}}_{\nu}  =&~ Y_{\nu}^{i1}\overline{L^{i}} \tilde{\varphi} \nu_{_{\mathtt{R}}}^{1}+ Y_{\nu}^{i2(3)}\overline{L^{i}} \tilde{\Phi} \nu_{_{\mathtt{R}}}^{2(3)}+ \frac{M_{\nu}^{pq}}{2}
\overline{(\nu_{_{\mathtt{R}}}^{p})^{c}}\nu_{_{\mathtt{R}}}^{q} +  \frac{Y_M^{p3}}{2} \overline{(\nu_{_{\mathtt{R}}}^{p})^{c}}\nu_{_{\mathtt{R}}}^{3} \chi+    \text{h.c.}.
\end{split}
\end{equation}
After symmetry breaking, we can write mass terms as,
\begin{equation}
- \mathcal{L_{M}}_{\nu} = \overline{\nu_{_{\mathtt{L}}}} \mathcal{M}_{D}^{\nu}\nu_{_{\mathtt{R}}} + \frac{1}{2} \overline{\nu_{_{\mathtt{R}}}^{c}} \mathcal{M}_{R} \overline{\nu_{_{\mathtt{R}}}}\, + \text{h.c.}.
\end{equation}
With this, the Majorana mass matrix can be written as,
\begin{align}\label{Eq:Majotana_mass_matrix}
-\mathcal{L_{M}}_{\nu} = \quad \frac{1}{2}
\begin{pmatrix}
\overline{\nu_{_{\mathtt{L}}}}  &  \overline{\nu_{_{\mathtt{R}}}^{c}}
\end{pmatrix} \begin{pmatrix}
0 &  \mathcal{M}_{D}^{\nu} \\
(\mathcal{M}_{D}^{\nu})^T & \mathcal{M}_{R}
\end{pmatrix} \begin{pmatrix}
\nu_{_{\mathtt{L}}}^c \\
\nu_{_{\mathtt{R}}}
\end{pmatrix} + \text{h.c.}\,.
\end{align}
After diagonalization of the above mass matrix, one can easily recover the seesaw mass formula, $m_{\nu} \simeq -\mathcal{M}_{D}^{\nu} \mathcal{M}_{R}^{-1} (\mathcal{M}_{D}^{\nu})^T$ and $m_{R} \simeq \mathcal{M}_{R}$, in the seesaw limit of $\mathcal{M}_{D}^{\nu} \ll \mathcal{M}_{R}$.

\subsection{Fermions coupling with $\mathbf{Z'}$}
As discussed above, there are no $Z'$-induced FCNCs due to our choice of fermions $U(1)_X$ charges and rotation matrices. This makes the $Z'$ couplings to fermions identical to those in the gauge basis, as given in Eq.~\eqref{Eq:VA_coup}. 
Using the charge assignments summarized in Table~\ref{Tab:parametertable1}, we compute the vector and axial-vector couplings of the proton and neutron using Eqs. \eqref{Eq:Proton_Neutron_Vector_coupnig_def}, and \eqref{Eq:Proton_Neutron_Axial_coupnig_def}. The resulting couplings are given by
\begin{equation}\label{Eq:Proton_Neutron_coup_sol1}
    a_{p}=a_{n} = 0.013\, X_{\nu} g_x,~~ C_p = 2 \left( \frac{\tilde{g}}{g_{x}}-X_{L} \right)\cos^2 \theta_{W} g_x \equiv 2 x g_x,~~ C_n = 0\,,
\end{equation}
where for convenience, to parametrize the vector coupling of the $Z'$ to the proton $C_p$, we have defined the parameter $x$,
\begin{equation}
    x = \left( \frac{\tilde{g}}{g_{x}} - X_{L} \right)\cos^2\theta_{W}\,.
\end{equation}
Using this parametrization the vector and axial-vector couplings of the other relevant 
fermions are presented in 
Table~\ref{Tab:Coupling1}.
\begin{table}[ht]
\centering
\renewcommand{\arraystretch}{2}
\setlength{\tabcolsep}{20pt} 
\small
\begin{tabular}{|c c c |}
\hline
Fields ($\psi$) & $C_{V}^{\psi}$ & $C_{A}^{\psi}$  \\ 
\hline
$u$ & $\frac{4}{3} xg_x$ & $0$ \\ 
$d$ & $-\frac{2}{3}x g_x$ & $0$  \\ 
$e$ & $-2x g_x $ & $0$ \\ 
\hline
$s$ & $-\frac{1}{6} ( 3 X_{\nu} + 4 x)g_x $ & $-\frac{1}{2}X_{\nu} g_x$  \\
$\mu$ & $-2x g_x$ & $0$ \\
\hline
\end{tabular}
\caption{Vector ($C_V^\psi$) and axial-vector ($C_A^\psi$) couplings of various fermions ($\psi=u,d,e,s,\mu$) to the $Z'$ boson, as relevant for the phenomenological analysis in this work.}
\label{Tab:Coupling1}
\end{table}
From Table~\ref{Tab:Coupling1}, one can see that the axial-vector couplings vanish for first generation quarks and leptons. Consequently, the axial-vector couplings of the proton and neutron in Eq.~\eqref{Eq:Proton_Neutron_coup_sol1}, arise solely from the strange quark contribution.
Furthermore, since the $Z'$ neutrino coupling vanishes, the $Z'$ decays only into an electron-positron pair, leading to ${\rm Br}(Z' \to e^{+} e^{-}) = 1$.
Other features and constraints on these couplings are:
\begin{itemize}
    \item Since $C_A^e = 0$, the model evades the constraints from atomic parity violation observables presented in Eqs.~\eqref{Eq:Cons_APV} and \eqref{Eq:Cons_APV_Moller}.
\item The vector couplings of the electron ($C_V^e$) and the proton ($C_p$) 
to the $Z'$ are both proportional to $x g_x$. 
Consequently, experimental constraints on these interactions can be 
translated into bounds on $|x g_x|$.
The most stringent upper limit arises from protophobic constraints 
derived from NA48, as discussed in Eq.~\eqref{Eq:Const_Protophobic}, 
which yield $|x g_x| \lesssim 1.25 \times 10^{-4}$. 
Additional constraints from the electron sector originate from the 
anomalous magnetic moments of the electron and muon -- see Eq.~\eqref{Eq:cons:gm2} -- implying
$\Delta a_e^\text{BSM}: \; |x g_x| \lesssim 1.59 \times 10^{-4}$ 
and 
$\Delta a_\mu^\text{BSM}: \; |x g_x| \lesssim 10^{-3}$. 
However, both bounds are weaker than the one coming from the protophobic requirement.
Electron-positron annihilation searches at KLOE 
(see Eq.~\eqref{Eq:Const_epen_scattering}) 
also impose limits on $|x g_x|$, although these are less restrictive 
than those from NA48 and $(g-2)_e$.
A lower bound on this product arises from the requirement that the $Z'$ 
decay products be absorbed in the dump, corresponding to the 
strong coupling regime of electron beam dump experiments 
(see Eq.~\eqref{Eq:Cons_Beam_dump}), which demands 
$|x g_x| \gtrsim 1.8 \times 10^{-5}$.
Therefore, the allowed range of this parameter is,
\begin{equation}
    1.8 \times 10^{-5} \lesssim |x g_x| \lesssim 1.25 \times 10^{-4} \, .
\end{equation}
\item Given that both $\Delta Q_{Z'}$ and $C_n$ vanish, the model evades the constraints from SINDRUM-I and neutron lead scattering presented in Eqs.~\eqref{Eq:Const_SINDRUM} and \eqref{Eq:Const_Neutron_lead}, respectively. Furthermore, with the $Z'$--neutrino coupling also vanishing, there are no constraints from neutrino scattering experiments. 
\item Vector coupling of the strange quark is proportional to $(3X_{\nu}+4x) g_x$, for which limits can be obtained from the KLOE-2 experiment search for $\phi \to \eta \, (Z' \to e^{+} e^{-})$. However, it only constrains $M_{Z'}$ above 30 MeV \cite{KLOE-2:2012lii}. Hence, the parameter $``X_{\nu}"$ remains unconstrained and can therefore be varied independently.
\end{itemize}


\section{Results}\label{Sec:results}

Having discussed the allowed range of couplings, we now examine the parameter space required to accommodate the ATOMKI anomaly.
Our analysis centers on the mixed vector--axial-vector scenario. In general, when both vector and axial-vector coupling are present, the decay width for on shell $Z'$ emission is the direct sum of the two contributions.
However, in our case, the vector and axial-vector couplings are of comparable magnitude. In this case, the axial-vector contribution typically dominates~\cite{Barducci:2022lqd}. Therefore, we adopt the fit for the pure axial-vector scenario presented in Ref.~\cite{Barducci:2022lqd}.
\begin{figure}[t!]
\begin{center}
\includegraphics[width=0.49\linewidth]{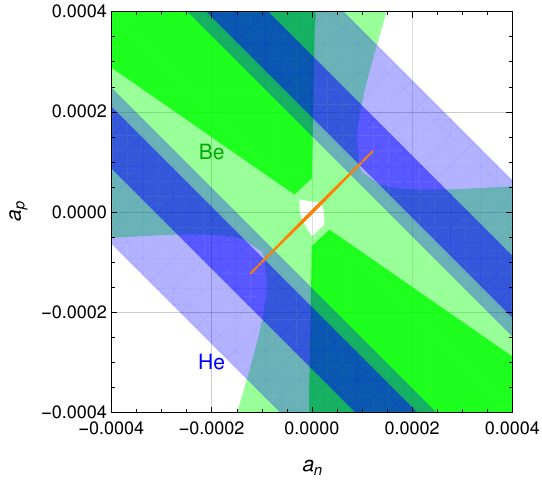}
\includegraphics[width=0.49\linewidth]{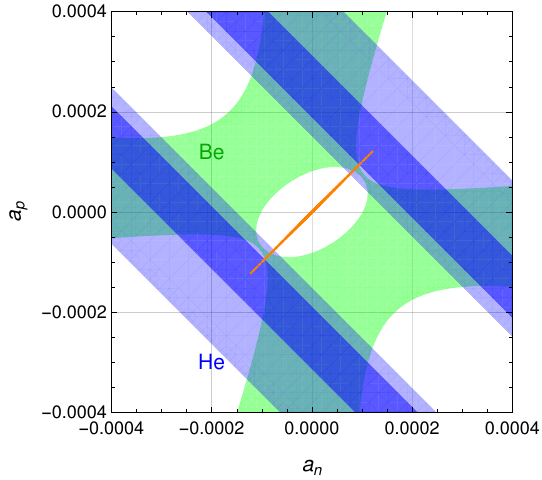}
\end{center}
\caption{The parameter space in the space in the ($a_n$-$a_p$) plane for two scenarios.  
\textbf{Left:} fit including only the $^8$Be (18.15 MeV) excited state.  
\textbf{Right:} simultaneous fit to both $^8$Be (18.15 MeV) and $^8$Be (17.64 MeV) excited states.  
Dark (light) shaded regions represent the $1\sigma$ ($2\sigma$) allowed ranges accommodating the combined $^8$Be (green) and $^4$He (blue) anomalies.
The orange points show the model predictions, which satisfy all experimental constraints discussed in Section~\ref{Sec:Exp_Constraints}.}
\label{FiG:Model}
\end{figure}
In Fig.~\ref{FiG:Model}, we present the allowed parameter space in the 
($a_n,a_p$) plane for two different scenarios. 
The left panel corresponds to the fit including only the 
$^{8}\text{Be}(18.15)$ excited state, whereas the right panel 
shows the case where both the $^{8}\text{Be}(18.15)$ and 
$^{8}\text{Be}(17.64)$ excited states are taken into account.
The darker (lighter) shaded regions correspond to the $1\sigma$ ($2\sigma$) 
ranges within which the $^{8}\text{Be}$ (green) and $^{4}\text{He}$ (blue) anomalies can be accommodated. The orange curve represents the model prediction obtained by varying $X_{\nu}\, g_x$ in the range $[0,\,10^{-2}]$.
In both cases, the model satisfies the $^{8}\text{Be}$ and $^{4}\text{He}$ anomaly within $2\sigma$ for a small region of the parameter space, as shown in Fig.~\ref{FiG:Model}.

It is worth noting that the charge assignments for different generations in Table~\ref{Tab:parametertable1} are not unique.
Since the gauge anomaly cancellation conditions depend only on the 
generation wise sums (and sums of squares or cubes) of the charges, 
any permutation of the fermion generation labels leaves the anomaly 
cancellation conditions unchanged.
In particular, the $U(1)_X$ charges of fermions belonging to the same 
SM representation, such as the lepton doublets 
$L^1, L^2,$ and $L^3$, or the quark singlets 
$d_{\mathtt{R}}^{1}, d_{\mathtt{R}}^{2},$ and $d_{\mathtt{R}}^{3}$, can be interchanged without affecting the 
anomaly free nature of the charge assignment.
Hence, by rearranging the charges in Table~\ref{Tab:parametertable1}, additional solutions can be obtained. However, it is challenging to simultaneously accommodate the beryllium and helium anomalies while satisfying all experimental constraints. Two such models are presented in the appendix \ref{App:Models_SINDRUMI}.

\section{Conclusion}\label{sec:conclusion}
The ATOMKI Collaboration has reported persistent anomalies in the internal pair creation decays of excited states in $^8$Be, $^4$He, and $^{12}$C nuclei, consistently pointing toward the on shell production of a light boson $  X  $ with mass around 17 MeV. 
In this work, we have investigated the possibility of explaining the 
ATOMKI anomalies within a class of gauged flavor specific chiral 
$U(1)_X$ extensions of the SM. 
Motivated by existing 
spin parity analyses and experimental constraints, we focused on spin-1 mediators with axial-vector or mixed vector--axial-vector 
couplings, which remain the most viable candidates for a simultaneous 
explanation of the observed anomalies.

We showed that constructing a consistent  framework 
that realizes such couplings is highly nontrivial due to the combined requirements of gauge anomaly cancellation and fermion mass generation.
In particular, we emphasized that in minimal single Higgs constructions, 
axial-vector couplings are suppressed in the light $Z'$ regime.
This suppression can be naturally avoided in a two Higgs doublet framework with flavor specific chiral $U(1)_X$ charge assignments.

Given the significant theoretical uncertainties associated with the 
nuclear matrix elements in the $^{12}\mathrm{C}$ transition, we focused 
primarily on the $^{8}\mathrm{Be}$ 
and $^{4}\mathrm{He}$ anomalies.
Within this setup, we systematically constructed anomaly-free charge assignments and demonstrated that the resulting models can generate the required axial-vector 
couplings to SM fermions.
We identified viable regions of parameter space where $^{8}\mathrm{Be}$ 
and $^{4}\mathrm{He}$ anomalies can be simultaneously explained, 
while satisfying constraints from atomic parity violation, beam dump 
experiments, anomalous magnetic moments, meson decays, neutrino 
scattering, and collider searches.

Our results highlight that flavor specific chiral $U(1)_X$ models with 
two Higgs doublets provide a consistent and predictive framework 
for realizing light axial-vector mediators. These models offer a 
promising avenue for explaining the ATOMKI anomalies and can be further 
tested in upcoming low-energy experiments. 
Future progress, particularly in improving nuclear matrix element 
calculations for $^{12}\mathrm{C}$ and independent experimental 
confirmation, will be crucial in clarifying the origin of these anomalies.

\acknowledgments
\noindent
The work of H.P. is supported by the Prime Minister Research Fellowship (ID: 0401969). A.B. is supported by the PhD grant UI/BD/154391/2023 from the Fundação para a Ciência e a Tecnologia (FCT, Portugal). The work of A.B. and F.R.J. is also supported  by the \href{https://doi.org/10.54499/UID/00777/2025}{FCT project UID/00777/2025}.

\appendix
\section{Additional Models satisfying ATOMKI anomaly and other experimental constraints but in conflict with SINDRUM-I}\label{App:Models_SINDRUMI}

In this appendix, we present alternative anomaly-free solutions obtained 
by permuting the $U(1)_X$ charge assignments introduced in 
Table~\ref{Tab:parametertable1}. The primary solution discussed in the 
main text is flavor universal within the quark and lepton doublet 
sectors.
Since the gauge-anomaly cancellation conditions depend on the generation wise sums and the sums of the squares 
and cubes of the $U(1)_X$ charges, the right handed quark singlet charges 
can be permuted to generate additional anomaly free solutions.
Two such configurations, denoted as Model~I and Model~II, are discussed below. While both models can accommodate the observed $^8\text{Be}$ and $^4\text{He}$ anomalies, the allowed parameter space is in tension with limits from the SINDRUM-I experiment at the 90\% confidence level.

\subsection{Model~I}

\begin{table}[ht]
\centering
\renewcommand{\arraystretch}{2}
\setlength{\tabcolsep}{20pt} 
\small
\begin{tabular}{|c c|c c|}
\hline
Fields & $U(1)_X$ & Fields & $U(1)_X$ \\ 
\hline
$Q^{1},Q^{2},Q^{3}$ & $-\frac{X_{L}}{3}$ & $L^{1},L^{2},L^{3}$ & $X_{L}$ \\ 
$u_{\mathtt{R}}^{1},u_{\mathtt{R}}^{2},u_{\mathtt{R}}^{3}$ & $\left(-\frac{4X_{L}}{3},~X_{\nu} - \frac{4X_{L}}{3},~-\frac{4X_{L}}{3} \right)$ & $e_{\mathtt{R}}^{1},e_{\mathtt{R}}^{2},e_{\mathtt{R}}^{3}$ & $(2X_{L},~2X_{L},~2X_{L}-X_{\nu})$ \\ 
$d_{\mathtt{R}}^{1},d_{\mathtt{R}}^{2},d_{\mathtt{R}}^{3}$ & $\left(-X_{\nu} + \frac{2X_{L}}{3},~\frac{2X_{L}}{3},~\frac{2X_{L}}{3} \right)$ & $\nu_{\mathtt{R}}^{1},\nu_{\mathtt{R}}^{2},\nu_{\mathtt{R}}^{3}$ & $(0,~0,~X_{\nu})$ \\ 
\hline
$\Phi$ & $-X_{L}$ & $\chi$ & $ -X_{\nu} $ \\
$\varphi$ & $X_{\nu} - X_{L}$ & & \\
\hline
\end{tabular}
\caption{Charges of the SM and BSM particles under the $U(1)_X$ gauge group for Model I.}
\label{Tab:parametertable2}
\end{table}
The $U(1)_X$ charge assignments of the particles in this model are presented in Table~\ref{Tab:parametertable2}. In this case, the right-handed down quark charges of the first and second generations are interchanged relative to those given in 
Table~\ref{Tab:parametertable1}.
The corresponding vector and axial-vector couplings with fermions can be computed from these charges using Eq.~\eqref{Eq:VA_coup}, as listed in Table~\ref{Tab:Coupling2}.
\begin{table}[ht]
\centering
\renewcommand{\arraystretch}{2}
\setlength{\tabcolsep}{20pt} 
\small
\begin{tabular}{|c c c |}
\hline
Fields ($\psi$) & $C_{V}^{\psi}$ & $C_{A}^{\psi}$  \\ 
\hline
$u$ & $\frac{2}{3}x g_x$ & $0$ \\ 
$d$ & $-\frac{1}{6}(3X_{\nu}+2x )g_x$ & $-\frac{1}{2}X_{\nu}g_x$  \\ 
$e$ & $-x g_x $ & $0$ \\ 
\hline
$s$ & $-\frac{1}{3} x g_x $ & $0$  \\
$\mu$ & $-x g_x$ & $0$ \\
\hline
\end{tabular}
\caption{Vector ($C_V^\psi$) and axial-vector ($C_A^\psi$) couplings of various fermions ($\psi=u,d,e,s,\mu$) to the $Z'$ boson for Model~I}
\label{Tab:Coupling2}
\end{table}
Here the parameter x is defined as, $x=2\left( \tilde{g}/g_{x} - X_{L} \right)\cos^2\theta_{W}\,.$
From these couplings, the effective proton and neutron couplings can be obtained using Eq.~\eqref{Eq:Proton_Neutron_Vector_coupnig_def} and Eq.~\eqref{Eq:Proton_Neutron_Axial_coupnig_def} as:
\begin{equation}\label{Eq:Proton_Neutron_coup_sol3}
a_{p}= 0.18\, X_{\nu} g_x,~~  a_{n} = -\, 0.45\, X_{\nu} g_x,~~ C_p = -\, 0.5 (X_{\nu}-2x) g_x,~~ C_n = -\, X_{\nu}g_x\,.
\end{equation}
Note that in this case, $a_p$ and $a_n$ differ from each other and, consequently, the slope of the line in the ($a_n,a_p$) plane deviates from that given in Eq.~\eqref{Eq:Proton_Neutron_coup_sol1}.
The electron axial-vector coupling is again zero, since the lepton charges remain identical to those in Table~\ref{Tab:parametertable1}. Consequently, there are no constraints from parity violation observables.
The electron vector coupling is $x g_x$ and its allowed range is bounded from above by the $(g-2)_e$ constraint and from below by beam-dump experiments: $3.6 \times 10^{-5}\lesssim \lvert x g_x\rvert \lesssim 3.2 \times 10^{-4}$. The proton vector coupling is proportional to $X_{\nu} - 2x$, and protophobic constraints impose the limit: $\lvert X_{\nu}-2x  \rvert g_x \lesssim 5\times 10^{-4} $.
Since $\Delta Q_{Z'} \neq 0$ in this case, a very strong additional constraint arises from the SINDRUM-I experiment: $|\Delta Q_{Z'}| = |X_{\nu} g_x| \lesssim 8.5 \times 10^{-5}$.
\begin{figure}[ht]
\begin{center}
\includegraphics[width=0.49\linewidth]{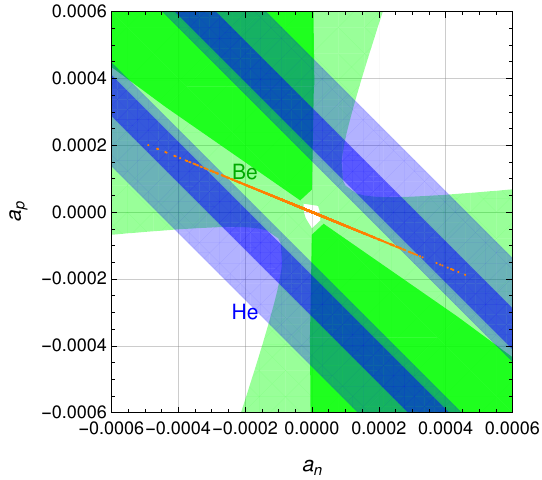}
\includegraphics[width=0.49\linewidth]{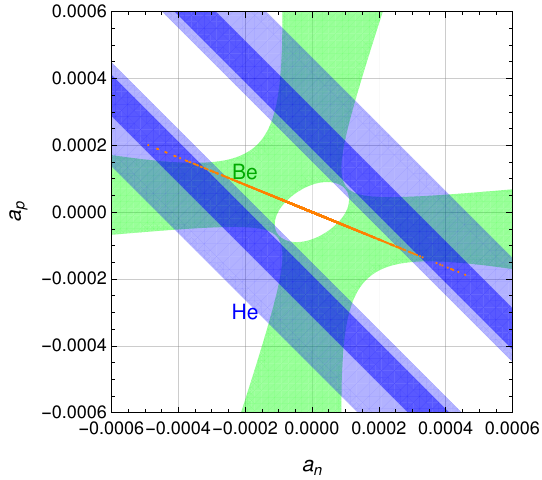}
\end{center}
\caption{Parameter space in the $(a_n,a_p)$ plane for Model I. Color coding is the same as in Fig.~\ref{FiG:Model}. The allowed regions that simultaneously accommodate both the $^4$He and $^8$Be anomalies are in tension with the SINDRUM-I limits at the $90\%$ confidence level. See text for details.}
\label{FiG:Model_1}
\end{figure}
In Fig.~\ref{FiG:Model_1}, we display the allowed parameter space in the 
($a_n,a_p$) plane. The left panel corresponds to the case where only the 
$^{8}\text{Be}(18.15)$ excited state is included, while the right panel 
shows the scenario in which both the $^{8}\text{Be}(18.15)$ and 
$^{8}\text{Be}(17.64)$ excited states are taken into account.
The model prediction is indicated by the orange points. In generating 
these points, we impose all experimental constraints except those 
from the SINDRUM-I experiment.
The dark (light) shaded region corresponds to the $1\sigma$ ($2\sigma$) 
allowed range in which the $^{8}\text{Be}$ (green) and 
$^{4}\text{He}$ (blue) anomalies can be accommodated.
In both scenarios, simultaneously explaining the $^{8}\text{Be}$ and 
$^{4}\text{He}$ anomalies requires $|X_{\nu} g_x| \gtrsim 10^{-4}$, which is 
in tension with the SINDRUM-I bound at $90\%$ C.L.

\subsection{Model~II}
Here we present another model obtained by rearranging the charges of the quark singlets. The corresponding $U(1)_X$ charge assignments are shown in Table~\ref{Tab:parametertable3}.
\begin{table}[ht]
\centering
\renewcommand{\arraystretch}{2}
\setlength{\tabcolsep}{20pt} 
\small
\begin{tabular}{|c c|c c|}
\hline
Fields & $U(1)_X$ & Fields & $U(1)_X$ \\ 
\hline
$Q^{1},Q^{2},Q^{3}$ & $-\frac{X_{L}}{3}$ & $L^{1},L^{2},L^{3}$ & $X_{L}$ \\ 
$u_{\mathtt{R}}^{1},u_{\mathtt{R}}^{2},u_{\mathtt{R}}^{3}$ & $\left(X_{\nu} - \frac{4X_{L}}{3},~-\frac{4X_{L}}{3} ,~-\frac{4X_{L}}{3} \right)$ & $e_{\mathtt{R}}^{1},e_{\mathtt{R}}^{2},e_{\mathtt{R}}^{3}$ & $(2X_{L},~2X_{L},~2X_{L}-X_{\nu})$ \\ 
$d_{\mathtt{R}}^{1},d_{\mathtt{R}}^{2},d_{\mathtt{R}}^{3}$ & $\left(\frac{2X_{L}}{3},~-X_{\nu} + \frac{2X_{L}}{3},~\frac{2X_{L}}{3} \right)$ & $\nu_{\mathtt{R}}^{1},\nu_{\mathtt{R}}^{2},\nu_{\mathtt{R}}^{3}$ & $(0,~0,~X_{\nu})$ \\ 
\hline
$\Phi$ & $-X_{L}$ & $\chi$ & $ -X_{\nu} $ \\
$\varphi$ & $X_{\nu} - X_{L}$ & & \\
\hline
\end{tabular}
\caption{Charges of the SM and BSM particles under the $U(1)_X$ gauge group for Model~II.}
\label{Tab:parametertable3}
\end{table}
In this case, the $U(1)_X$ charges assigned to the right-handed up quarks of the first and second generations are interchanged with respect to the assignments in Table~\ref{Tab:parametertable1}.
Using the $U(1)_X$ charges given in Table~\ref{Tab:parametertable1}, the vector and axial-vector couplings can be computed from Eq.~\eqref{Eq:VA_coup}. The resulting couplings are presented in Table~\ref{Tab:Coupling3}.
\begin{table}[ht]
\centering
\renewcommand{\arraystretch}{2}
\setlength{\tabcolsep}{20pt} 
\small
\begin{tabular}{|c c c |}
\hline
Fields ($\psi$) & $C_{V}^{\psi}$ & $C_{A}^{\psi}$  \\ 
\hline
$u$ & $\frac{1}{6}(3X_{\nu}+4x) g_x$ & $\frac{1}{2}X_{\nu}g_x$ \\ 
$d$ & $-\frac{1}{3}x g_x$ & $0$  \\ 
$e$ & $-x g_x $ & $0$ \\ 
\hline
$s$ & $-\frac{1}{6} ( 3 X_{\nu} + 2 x)g_x $ & $-\frac{1}{2}x g_x$  \\
$\mu$ & $-x g_x$ & $0$ \\
\hline
\end{tabular}
\caption{Vector ($C_V^\psi$) and axial-vector ($C_A^\psi$) couplings of various fermions ($\psi=u,d,e,s,\mu$) to the $Z'$ boson for Model~II}
\label{Tab:Coupling3}
\end{table}
, where $x=2\left( \tilde{g}/g_{x} - X_{L} \right)\cos^2\theta_{W}\,.$ 
With this charge assignment, the proton and neutron couplings are given by,
\begin{equation}\label{Eq:Proton_Neutron_coup_sol2}
a_{p}= 0.46\, X_{\nu} g_x,~~  a_{n} =- 0.17\, X_{\nu} g_x,~~ C_p = (X_{\nu}+x) g_x,~~ C_n = 0.5\,X_{\nu}g_x\,.
\end{equation}
The magnitude of the electron vector coupling remains the same as in Model I. Consequently, the allowed range for this coupling is identical to that in Model I:
$3.6 \times 10^{-5}\lesssim \lvert x g_x\rvert \lesssim 3.2 \times 10^{-4}$.
The constraint from the proton vector coupling is, $\lvert X_{\nu}+x  \rvert g_x \lesssim 2.5\times 10^{-4} $.
Additionally, in this model $\Delta Q_{Z'} = X_{\nu} g_x$. Consequently, the SINDRUM-I bounds impose similarly severe constraints as in Model I.
\begin{figure}[ht]
\begin{center}
\includegraphics[width=0.49\linewidth]{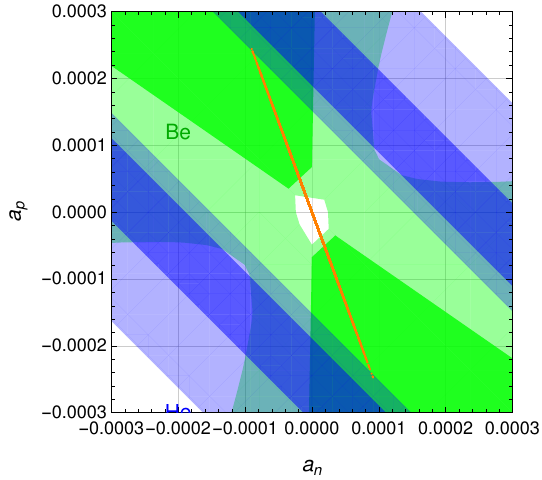}
\includegraphics[width=0.49\linewidth]{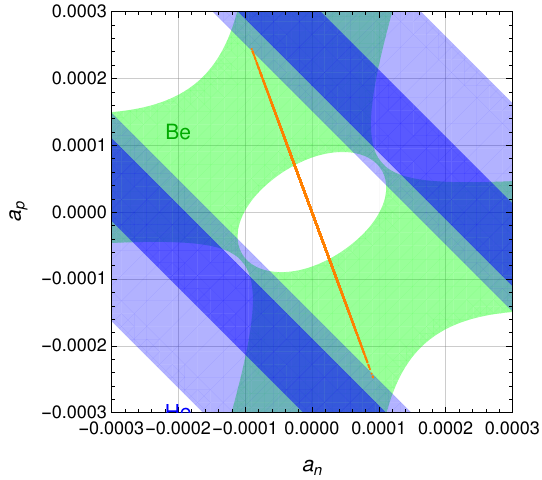}
\end{center}
\caption{Parameter space in the $(a_n,a_p)$ plane for Model II. Color coding is the same as in Fig.~\ref{FiG:Model}. The allowed regions that simultaneously accommodate both the $^4$He and $^8$Be anomalies are in tension with the SINDRUM-I limits at the $90\%$ confidence level. See text for details.}
\label{FiG:Model_2}
\end{figure}

In Fig.~\ref{FiG:Model_2}, we illustrate the allowed parameter space in the $(a_n,a_p)$ plane for Model~II. The left panel shows the results when only the $^8$Be (18.15 MeV) excited state is included in the fit, whereas the right panel incorporates both the $^8$Be (18.15 MeV) and $^8$Be (17.64 MeV) excited states. The orange points represent the model predictions. These points are obtained by applying all experimental constraints except the SINDRUM-I bound. The dark-shaded (light-shaded) regions indicate the $1\sigma$ ($2\sigma$) allowed parameter space that simultaneously accommodates the $^8$Be (green) and $^4$He (blue) anomalies. Both anomalies can be satisfied only in a very small region of parameter space, as illustrated in Fig.~\ref{FiG:Model_2} for both scenarios. Furthermore, accommodating both anomalies simultaneously requires $|X_{\nu} g_x|$ of order $\mathcal{O}(10^{-4})$, which is in conflict with the SINDRUM-I constraint at the $90\%$ confidence level.

\bibliographystyle{utphys}
\bibliography{bibliography}

\end{document}